\documentclass[11pt,nofootinbib]{revtex4-1}
\pdfoutput=1
\usepackage{graphicx}
\usepackage{amssymb}
\usepackage[english]{babel}
\usepackage{psfrag}
\newcommand{\bq}{\begin{equation}}
\newcommand{\eq}{\end{equation}}
\newcommand{\ba}{\begin{eqnarray}}
\newcommand{\ea}{\end{eqnarray}}
\usepackage{color}
\usepackage{hyperref}
\usepackage[]{indentfirst}
\usepackage[]{amsfonts}
\usepackage{amsmath}
\usepackage[top=1.25in, bottom=1.25in, left=1.25in, right=1.25in]{geometry}
\usepackage{bbold}
\usepackage{placeins}
\pdfoutput=1
\begin{document}

\title{Rock-paper-scissors played within competing domains in predator-prey games}
\author{Darka Labavi\'{c}$^1$, and Hildegard Meyer-Ortmanns$^1$}
\address{$^1$Physics and Earth Sciences, Jacobs University Bremen, P.O.Box 750561, 28725 Bremen, Germany}
%\eads{\mailto{d.labavic@jacobs-university.de},  \mailto{h.ortmanns@jacobs-university.de}}

\begin{abstract}
We consider $(N,r)$ games of prey and predation  with $N$ species and $r < N$ prey and predators, acting in a cyclic way. Further basic reactions include reproduction, decay and diffusion over a one- or two-dimensional regular grid, without a hard constraint on the occupation number per site, so in a ``bosonic" implementation. For special combinations of $N$ and $r$ and appropriate parameter choices we observe games within games, that is different coexisting games, depending on the spatial resolution. As concrete and simplest example we analyze the (6,3) game. Once the players segregate from a random initial distribution, domains emerge, which effectively play a (2,1)-game on the coarse scale of domain diameters, while agents inside the domains  play (3,1) (rock-paper-scissors), leading to spiral formation with species chasing each other. As the (2,1)-game has a winner in the end, the coexistence of domains is transient, while agents inside the remaining domain coexist, until demographic fluctuations lead to extinction of all but one species in the very end. This means that we observe a dynamical generation of multiple space and time scales with emerging re-organization of players upon segregation, starting from a simple set of rules on the smallest scale (the grid constant) and changed rules from the coarser perspective.  These observations are based on Gillespie simulations. We discuss the deterministic limit derived from a van Kampen expansion. In this limit we perform a linear stability analysis and numerically integrate the resulting equations. The linear stability analysis  predicts the number of forming domains, their composition in terms of species; it also explains the instability of interfaces between domains, which drives their extinction; spiral patterns are identified as motion along heteroclinic cycles. The numerical solutions reproduce the observed patterns of the Gillespie simulations including even extinction events, so that the mean-field analysis here is very conclusive, which is due to the specific implementation of rules.
\end{abstract}

\maketitle

\section{Introduction}
Evolutionary game theory is used on different levels of biological systems, ranging from the genetic level to ecological systems. The language of game theory allows to address basic questions of ecology, related to the emergence of cooperation and biodiversity, as well as the formation of coalitions with applications to social systems. Darwinian dynamics, in particular frequency-dependent selection, can be formulated in terms of game-theoretic arguments \cite{nowak}. The formation of dynamical patterns is considered as one of the most important promoters of biodiversity \cite{may,levin,durret,hassel}. Here we consider games of competition, where the competition is realized as predation among $N$ species, where each species preys on $r$ others in a cyclic way. A subclass of these $(N,r)$-games are cyclic games, that is $(N,1)$ with $(3,1)$ being the famous rock-paper-scissors game. An extensive overview on cyclic games is given in \cite{szabo,perc}. The $(3,1)$-game has been studied in various extensions (spatial, reproduction and deletion, swapping or diffusion, mutation). One of the first studies of a $(3,1)$-game without spatial assignment, but in a deterministic and stochastic realization revealed that fluctuations due to a finite number of agents  can drastically alter the mean-field predictions, including an estimate of the extinction probabilities at a given time \cite{reichen1}. This model was extended to include a spatial grid in \cite{reichen2}, where the role of stochastic fluctuations and spatial diffusion was analyzed both numerically and analytically. The influence of species mobility on species diversity was studied in \cite{reichen3}, pattern formation close to a bifurcation point was the topic of \cite{reichen4}, see also \cite{reichen5} and the impact of asymmetric interactions was considered in \cite{reichen6}.

An extension to four species, first without spatial assignment, shows interesting new features as compared to $(3,1)$: Already in the deterministic limit the trajectories show a variety of possible orbits, and from a certain conserved quantity the late-time behavior can be extrapolated \cite{durney}. The four species can form alliance pairs similarly to the Game of Bridge \cite{case}. Under stochastic evolution various extinction scenarios and the competition of the surviving set of species can be analyzed \cite{28}. Domains and their separating interfaces were studied in \cite{arXiv:1205.4914}. $(4,1)$ cyclic games on a spatial grid were the topic in \cite{szabosznaider,luetz}. A phase transition as a function of the concentration of vacant sites is identified between a phase of four coexisting species and a phase with two neutral species that protect each other and extend their domain over the grid. For an extension of this model to long-range selection see \cite{hua}.

In this paper we focus on  the $(6,3)$-game, including both spiral formation inside domains and domain formation. It is a special case of $(N,r)$-games, which were considered for $N\ge3$ and $r\ge1$ by \cite{m1,m2} and more recently by \cite{m3}. The authors of \cite{m1,m2} were the first to notice that for certain combinations of N and r one observes the coexistence of both spiral formation and domain formation. However, it should be noticed that our set of reactions, even if we specialize $(N,r)$ to the $(3,1)$-game, is similar, but not identical with the versions, considered in \cite{m1,m2,m3}  or in \cite{reichen1}-\cite{reichen6}. The seemingly minor difference refers to the implementation of an upper threshold to the occupation number of single sites (set to 1 or a finite fixed number), while we use a ``bosonic" version. We introduce a dynamical threshold, realized via deletion reactions, so that we need not explicitly restrict the occupation number per site. Due to this difference, the bifurcation structure of the mean-field equations is changed.

The reason why we are interested in the particular combination of $N=6$ and $r=3$ is primarily motivated by two theoretical aspects rather than by concrete applications. As to the first aspect, this game is one of the simplest examples of ``games within games" in the sense that the domains effectively play a $(2,1)$-game as transient dynamics on a coarse scale (the scale of the domain diameter), while the actors inside the domains  play a $(3,1)$-game on the grid scale. Finally, one of the domains gets extinct along with all its actors. As such, this game provides a simple, yet non-trivial example for a mechanism that may be relevant for evolution: In our case, due to the spatial segregation of species, the structural complexity of the system increases in the form of patterns of who is chasing whom, appearing as long-living transients, along with a seemingly change of the rules of the game that is played between the competing domains on the coarse scale, while the rules, which individuals use on the elementary grid sites, are not changed at all. As outlined by Goldenfeld and Woese \cite{goldenfeldwoese}, it is typical for processes in evolution, in particular  in ecology, that ``the governing rules are themselves changed", as the system evolves in time and the rules depend on the state. In our example it is spatial segregation, which allows for a change of rules from a coarse perspective, as we shall see.

As to the second aspect, an interesting feature of such an arrangement is the multitude of time and spatial scales that are dynamically generated. Concretely in the $(6,3)$-game the largest of the reaction/diffusion rates sets the basic time unit. When the species segregate and form domains, the next scale is generated: it is the time it takes the two domains to form until both cover the two-dimensional grid or the one-dimensional chain. The domains are not static, but play the $(2,1)$-game that has a winner in the end. So the extinction time of one of the domains sets the third scale. A single domain then survives, including the moving spirals from the remaining $(3,1)$-game inside the domain. The transients can last very long, depending on the interaction rates and the system size. In the very end, however, in a stochastic realization as well as due to the finite accuracy in the numerical solutions even in the mean-field description, only one out of the three species will survive, and the extinction of the other two species sets the fourth scale. Along with these events, spatial scales emerge, ranging from the basic lattice constant to the radii of spirals and the extension of the domains.

One of the challenges is to explore which of the observed features in the Gillespie simulations can be predicted analytically. We shall study the predictions on the mean-field level, which is rather conclusive in our ultralocal implementation of reactions and reproduces the results of the Gillespie simulations quite well, since fluctuations turn out to play a minor role for pattern formation. The deterministic equations are derived as the lowest order of a van Kampen expansion. The eigenvalues of the Jacobian are conclusive for the number of surviving species in a stable state, the composition of the domains, and transient behavior, which is observed in the Gillespie simulations. The mean-field equations, including the diffusion term, will be integrated numerically and compared to the results of the Gillespie simulations.

The paper is organized as follows. In section~\ref{sec_reactions} we present the model in terms of basic reactions and the corresponding master equation. For generic $(N,r)$ games we summarize in section~\ref{sec_vankampen} the derivation of the mean-field equations from a van Kampen expansion, followed by a stability analysis via the Jacobian  with and without spatial dependence for the specific $(6,3)$ game, and a derivation of the numerical solutions of the mean-field equations in section~\ref{sec_Jacobian}. In section~\ref{sec_numerical} we present our results from the Gillespie simulations in comparison to the mean-field results. Section~\ref{sec_conclusions} summarizes our conclusions and gives an outlook to further challenges related to this class of games. For comparison, the supplementary material contains a detailed stability analysis for the $(3,1)$-game with spiral formation and the $(3,2)$-game with domain formation, as well as the numerical solutions of the mean-field equations and the corresponding Gillespie simulations.

\section{Reactions and Master Equation}\label{sec_reactions}
We start with the simplest set of reactions that represent predation between individuals of different species, followed by reproduction, deletion
\begin{eqnarray}\label{eq:rec_sys}
	X_{\alpha,i}\, +\, X_{\beta,i} & \overset{k_{\alpha\beta}/V}{\longrightarrow} & X_{\alpha,i}  \label{pred}\\
	X_{\alpha,i}\,  & \overset{r_{
\alpha,i}}{\longrightarrow} & 2 X_{\alpha,i} \label{repr} \\
	2X_{\alpha,i}\,& \overset{p_{\alpha}/V}{\longrightarrow} & X_{\alpha,i}  \label{anih}
\end{eqnarray}	
and finally diffusion
\begin{eqnarray}\label{diff}
	X_{\alpha,i}\, & \overset{D_{\alpha}/h^2}{\longrightarrow} & X_{\alpha,j}.
\end{eqnarray}
$X_{\alpha,i}$ represents an individual of species $\alpha$ at lattice site $i$, while the total number of individuals of species $\alpha$ at site $i$ will be denoted with $n_{\alpha,i}$. (We use small characters $n$ for convenience, although the meaning of $n$ is not a density, but the actual occupation number of a certain species at a certain site.) In view of applications to ecological systems, each lattice site stands  for a patch housing a subpopulation of a metapopulation, where the patch is not further spatially resolved.  Eq.~(\ref{pred}) represents the predation of species $\alpha$ on species $\beta$ with rate $k_{\alpha\beta}/V$, where the parameter $V$ does not stand for the physical volume, but parameterizes the distance from the deterministic limit in the following way: According to our set of reactions, larger values of $V$ lead to higher occupation numbers $n_{\alpha,i}$ of species $\alpha$ at sites $i$, since predation and deletion events are rescaled with a factor $1/V$, and therefore to a larger total rate. The fluctuations in occupation numbers, realized via the Gillespie algorithm, are independent of $V$ or the occupation numbers of sites, since only relative rates enter the probabilities for a certain reaction to happen. Therefore the size of the fluctuations relative to the absolute occupation numbers or to the overall $V$ gets reduced for large V, that is, in the deterministic limit. Predation is schematically described in figure~\ref{(6,3)}.

Eq.~(\ref{repr}) represents reproduction events with rate $r_{\alpha}$, and Eq.~(\ref{anih}) stands for death processes of species $\alpha$ with rate $p_{\alpha}/V$. Death processes are needed to compensate for the reproduction events, since we do not impose any restriction on the number of individuals that can occupy lattice sites. Here we should remark why we implement death processes in the form of Eq.~\ref{anih} rather than simpler as $X_{\alpha,i} \overset{p_{\alpha}}{\longrightarrow} \oslash$. The latter choice could be absorbed in a term $(\rho-\gamma)\phi_i\equiv \tilde{\rho}\phi_i$ in the mean-field equation (\ref{eq:pde}) below with uniform couplings $\rho$ and $\gamma$. This choice would not lead to a stable coexistence-fixed point \cite{josef} and therefore not to the desired feature of games within games\footnote{For the $(6,3)$-game we would have 40 fixed points, the sign of the eigenvalues would then only depend on the sign of the parameter $\tilde{\rho}$. At $\tilde{\rho}=0$ all fixed points collide and exchange stability through a multiple transcritical bifurcation. For $\tilde{\rho}>0$ (the only case of interest), the system has no stable fixed points, and the numerical integration of the differential equations diverges. (Similarly for the (3,1)-game, for $\tilde{\rho}>0$, the trivial fixed point with zero species is always an unstable node, while the coexistence fixed point is always a saddle.)}.

The species diffuse within  a two-dimensional lattice, which we reduce to one dimension for simplicity if we analyze the behavior in more detail. We assume that there can be more than one individual of one or more species at each lattice site. Individuals perform a random walk on the lattice with rate $D_{\alpha}/h^d$, where $D_{\alpha}$ is the diffusion constant, $h$ the lattice constant and $d$ the dimension of the grid. Diffusion is described by Eq.~(\ref{diff}), where $i$ represents the site from which an individual hops, and $j$ is one of the neighboring sites to which it hops. It should be noticed that diffusion is the only place, which leads to a spatial dependence of the results, since apart from diffusion, species interact on-site, that is, within their patch.

In summary, the main differences to other related work such as references \cite{reichen1,reichen2,reichen3,reichen4,reichen5,reichen6,durney,case, 28,m1,m2,m3} are the ultralocal implementation of prey and predation, no swapping, no mutations as considered in \cite{mobilia1,mobilia2}, and a bosonic version with a dynamically ensured finite occupation number of sites. Even if qualitatively similar patterns like spirals or domains are generated in all these versions, the bifurcation diagram, that is, the stability properties and the mode of transition from one to another regime depend on the specific implementation.

We can now write a master equation for the probability of finding $\{n\}$ particles at time $t$ in the system for reaction and diffusion processes, where $\{n\}$ stands for $(n_{1,1},...,n_{N,L^d})$  and $N$ is the number of species, $L^d$ the number of sites.
\begin{eqnarray}\label{eq:me_reac}
 %\hspace*{-3cm}
\frac{\partial P^{reac} \left( \left\{ n \right\};t \right)}{\partial t} &=&
	 \underset{i}{\sum} \left\{
	 \underset{\alpha,\beta}{\sum} \frac{k_{\alpha\beta}}{V} \left[
	  n_{\alpha,i}\left(n_{\beta,i}+1\right)P\left(n_{\alpha,i},n_{\beta,i}+1,...;t \right)
	   - n_{\alpha,i}n_{\beta,i}P\left(\{n\};t \right) \right] \right.\nonumber \\
	  &+&\left. \underset{\alpha}{\sum}  \frac{p_{\alpha}}{V} \left[
	  \left( n_{\alpha,i}+1 \right) n_{\alpha,i} P\left( ...,n_{\alpha,i}+1,...;t \right)
	  - n_{\alpha,i}\left( n_{\alpha,i}-1 \right) P(\{n\};t) \right] \right. \nonumber \\
       &+& \left. \underset{\alpha}{\sum}  r_{\alpha} \left[
       (n_{\alpha,i}-1)P(n_{\alpha,i}-1,...;t) - n_{\alpha,i}P(\{n\};t) \right]
	  \right\}
\end{eqnarray}
\begin{eqnarray}\label{eq:me_diff}
%\hspace*{-3cm}
	\frac{\partial P^{diff} \left( \left\{ n \right\};t \right)}{\partial t} &=& \underset{\alpha}{\sum} \frac{D_\alpha}{h^2} \underset{\left\langle i,j \right\rangle}{\sum} \left[ (n_{\alpha,i}+1)P(...,n_{\alpha,i}+1,n_{\alpha,j}-1,...;t)-n_{\alpha,i}P(\{n\};t)\right. \nonumber\\
	&+&\left. (n_{\alpha,j}+1)P(...,n_{\alpha,i}-1,n_{\alpha,j}+1,...;t)-n_{\alpha,j}P(\{n\};t)\right]
\end{eqnarray}
with $n_{\alpha,i}\ge 1$ for all $\alpha,i$, and
\begin{equation}\label{eq7}
	\partial_tP = \partial_tP^{reac}+\partial_tP^{diff}.
\end{equation}

As uniform (with respect to the grid) random initial conditions we assume a Poissonian distribution on each site $i$
\begin{equation}
	P \left( \{n\} ;0 \right)=\underset{\alpha,i}{\prod}\left( \frac{\overline{n}^{n_{\alpha,i}}_{\alpha,0}}{n_{\alpha,i}!} e^{-\overline{n}_{\alpha,0}} \right),
\end{equation}
where $\overline{n}_{\alpha,0}$ is the mean initial number of individuals of species $\alpha$ per site.

\section{Derivation of the mean-field equations}\label{sec_vankampen}
The master equation is continuous in time and discrete in space.  The diffusion term is included as a random walk. Next one takes the continuum limit in space, in which the random walk part leads to the usual diffusion term in the partial differential equation (pde) for the concentrations $\varphi(\vec{x},t)\equiv n_\alpha(\vec{x},t)/V$. The mean-field equations can then be derived by calculating the equations of motion for the first moments $\langle n_\alpha(\vec{x},t)\rangle$ from the master equation, where the average is defined as $\langle n_\alpha(\vec{x},t)\rangle=\sum n_\alpha(\vec{x},t)P(\{n_\alpha(\vec{x},t)\})$ with $P(\{n_\alpha(\vec{x},t)\})$ being a solution of the master equation, and factorizing higher moments in terms of first-order moments.

Alternatively,  we insert the ansatz for the van Kampen expansion according to $n_\alpha=V\varphi_\alpha + \sqrt{V}\eta_\alpha$ in the reaction part. To leading order in $V$ we obtain the deterministic pde for the concentrations of the reaction part. Combined with the diffusion part this leads to the full pde that is given as Eq.~\ref{eq:pde} in the next section. While this leading order then corresponds to the mean-field level, the next-to-leading order leads a Fokker-Planck equation with associated Langevin equation, from which one can determine the power spectrum of fluctuations. In our realization, the visible patterns are not fluctuation-induced, differently from noise-induced fluctuations as considered in \cite{goldenbutler}. Therefore our power spectrum of fluctuations is buried under the dominating spectrum that corresponds to patterns from the mean-field level. Therefore we do not further pursue the van Kampen expansion here\footnote{For details of a possible derivation of the mean-field equations we refer to \cite{darkathesis}; however, there we derived the mean-field equations via a longer detour towards a field theoretic formulation, where we read off the mean-field equations as leading order of a van Kampen expansion, not applied to the master equation, but to a Lagrangian that appears in the path integral derived from the master equations, in analogy to the derivation in \cite{goldenbutler}.}.

\section{Stability analysis of the mean field equations and their solutions}\label{sec_Jacobian}
We perform a linear stability analysis of the mean field equations by finding the fixed points of the system of partial differential equations
\begin{equation}\label{eq:pde}
	\partial_t\varphi_\alpha =
	D_\alpha\nabla^2\varphi_\alpha+
	r_\alpha\varphi_\alpha-
	p_\alpha\varphi_\alpha^2 -
	\underset{\beta}{\sum}k_{\beta\alpha}\varphi_\alpha\varphi_\beta,
\end{equation}
with $\phi_\alpha$ the concentration of species $\alpha$,
by setting $\partial_t\varphi_\alpha =D_\alpha\nabla^2\varphi_\alpha=0$. We will focus on the system with homogeneous parameters $r_\alpha=\rho$, $k_{\alpha\beta}=\kappa$ if species $\alpha$ preys on $\beta$ and 0 otherwise, $p_\alpha=\gamma$, and $D_\alpha=\delta$, $\forall\alpha,\beta\in\{1,...,N\}$ and consider the special case of the (6,3)-game. After finding the fixed points, we look at the eigenvalues of the Jacobian $J$ of the system~(\ref{eq:pde}) to determine the stability of the fixed points. We then extend our analysis to a spatial component by analyzing a linearized system in Fourier space, with Jacobian~\cite{cianci}
\begin{equation}
J^{SP}=J+\underline{D}\tilde{\Delta},
\end{equation}
where $\tilde{\Delta}=-k^2$ is the Fourier transform of the Laplacian and $\underline{D}$ is the diffusion matrix evaluated at a given fixed point. In our case, the diffusion matrix is a diagonal matrix $\underline{D}=\delta\mathbb{1}$. This leads to a dependence of the stability of the fixed points on diffusion. In the following we focus on the special case to be considered.

\subsection{Stability analysis and numerical integration for the (6,3)-game}
{\bf Stability analysis of the (6,3)-game.}
The (6,3)-game is given by the system of mean field equations:
\begin{eqnarray}\label{eq:MF(6,3)}
\frac{\partial\varphi_1}{\partial t} & = & \delta\nabla^2\varphi_1 + \rho\varphi_1 - \gamma\varphi_1^2 - \kappa\varphi_1(\varphi_4+\varphi_5+\varphi_6) \nonumber \\
\frac{\partial\varphi_2}{\partial t} & = & \delta\nabla^2\varphi_2 + \rho\varphi_2 - \gamma\varphi_2^2 - \kappa\varphi_2(\varphi_5+\varphi_6+\varphi_1) \nonumber \\
\frac{\partial\varphi_3}{\partial t} & = & \delta\nabla^2\varphi_3 + \rho\varphi_3 - \gamma\varphi_3^2 - \kappa\varphi_3(\varphi_6+\varphi_1+\varphi_2) \nonumber\\
\frac{\partial\varphi_4}{\partial t} & = & \delta\nabla^2\varphi_4 + \rho\varphi_4 - \gamma\varphi_4^2 - \kappa\varphi_4(\varphi_1+\varphi_2+\varphi_3) \nonumber\\
\frac{\partial\varphi_5}{\partial t} & = & \delta\nabla^2\varphi_5 + \rho\varphi_5 - \gamma\varphi_5^2 - \kappa\varphi_5(\varphi_2+\varphi_3+\varphi_4) \nonumber\\
\frac{\partial\varphi_6}{\partial t} & = & \delta\nabla^2\varphi_6 + \rho\varphi_6 - \gamma\varphi_6^2 - \kappa\varphi_6(\varphi_3+\varphi_4+\varphi_5). \nonumber\\
\end{eqnarray}
In total there are 64 different fixed points $FP_1$ to $FP_{64}$, of which some have the same set of eigenvalues and differ only by a permutation of the fixed-point coordinates of the eigenvalues,
so that we can sort all fixed points in 12 groups $FP^1$-$FP^{12}$: for example, the  fixed points $(\rho/(\gamma+\kappa),0,\rho/(\gamma+\kappa),0,\rho/(\gamma+\kappa),0)$ and $(0,\rho/(\gamma+\kappa),0,\rho/(\gamma+\kappa),0,\rho/(\gamma+\kappa))$ are in the same group $FP^3$. We will refer to all fixed points by the number of the group they belong to, that is to $FP^1$ to $FP^{12}$,  instead of $FP_1$ to $FP_{64}$.\\
The zero-fixed point $FP^1$, where all components are equal to zero, with all eigenvalues equal to $\rho$ for $\delta=0$, and equal to $\rho-\delta k^2$ for $\delta\neq0$, is unstable for a system without spatial assignment, while it can become stable for a spatial system if $\rho<\delta k^2$, as in the cases of the (3,1) and (3,2) games, which are discussed in detail in the supplementary material. In the coexistence-fixed point $FP^2$, all components are equal to $\rho/(\gamma+3 \kappa)$. It is stable for $\kappa/\gamma<1$, three of the eigenvalues are always negative, the first one being $-\rho$ and an the second and third one equal to $-\rho\gamma/(\gamma+\kappa)$, two are complex conjugates $-\rho(\gamma-\kappa\pm i \sqrt{3}\kappa)/(\gamma+3\kappa)$, and the last one is real $-\rho(\gamma-\kappa)/(\gamma+3\kappa)$. At $\kappa/\gamma=1$, $FP^2$ becomes a saddle, three of the six eigenvalues change sign, complex conjugates change sign of their real part, so a Hopf bifurcation occurs, and the direction corresponding to the last eigenvalue becomes unstable.\\
Other fixed points include the survival of one species ($FP^4$), two species (for both $FP^5$ and $FP^6$), three (for $FP^7$ and $FP^8$), four ( for $FP^9$, $FP^{10}$ and $FP^{11}$), and five species (for $FP^{12}$). All fixed points $FP^4$ to $FP^{12}$ are always saddles in the case of $\delta=0$.\\
For $\delta\neq0$ all eigenvalues get a $(-\delta k^2)$-term, which can extend the stability regime in the parameter space, as long as $k\neq0$, and lead to the coexistence of stable fixed points, which cannot be found for $\delta=0$.

\begin{figure}[ht]
	\begin{center}
		\includegraphics[width=5cm]{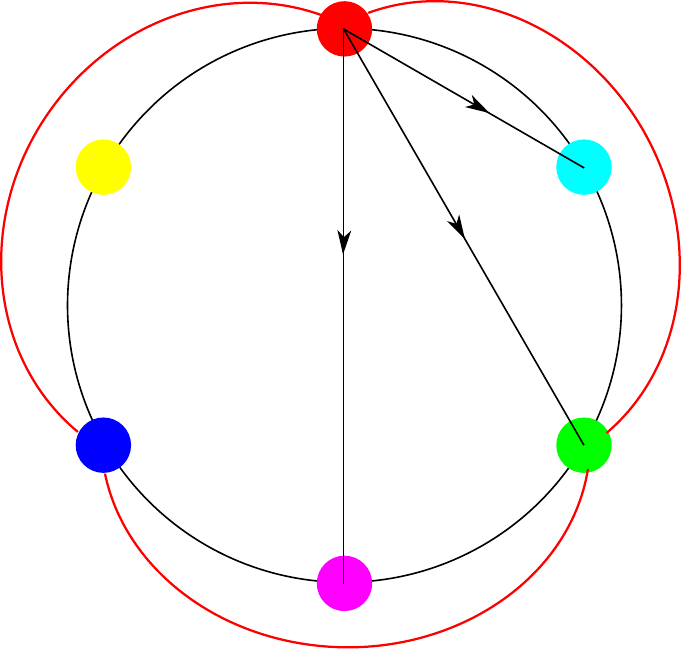}
		\end{center}
	\caption{Diagram of a (6,3)-game. Colors represent species, each preys on three other species in clockwise direction, shown only for the red species by black arrows. Red lines connect species, which form one domain (red, green, and blue), the other three species (cyan, magenta, and yellow) form the second domain. Each species (like the red one) preys on only one species from its own domain (green), and on two species from the other domain (cyan and magenta), this way eliminating all predators of the third species (blue) from the domains. These rules are characteristic for all games, in which a domain forms of three species playing the (3,1)-game. Colors in this scheme will be used throughout the paper to represent species one (red) to six (yellow). }\label{(6,3)}
\end{figure}

In view of pattern formation we shall distinguish three regimes. Before we go into detail, let us first give an overview of the sequence of events, if we vary the bifurcation parameters $\kappa$ and $\gamma$ as $\kappa/\gamma$. These are events, which we see both in the Gillespie simulations and the numerical integration of the mean-field equations in space, described as a finite grid upon integration.
\begin{itemize}
\item  $\kappa/\gamma<1$: the first regime with $\kappa/\gamma$ smaller than its value at the first Hopf bifurcation at $\kappa/\gamma=1$, where the 6-species coexistence fixed point becomes unstable. As long as this fixed point is stable, we see no patterns, as the system converges at each site of the grid to the 6-species fixed point without dominance of any species, so that the uniform color is gray.
\item $1<\kappa/\gamma<2$: the second regime with $\kappa/\gamma$ chosen between the first and second Hopf bifurcations, where the second one happens at $\kappa/\gamma=2$ for the $FP^3$ fixed points. When  $\kappa/\gamma=1$ is approached from below, that is from $\kappa/\gamma<1$, two fixed points, belonging to the $FP^3$-group, become stable through a transcritical bifurcation until $\kappa/\gamma=2$, where they become unstable through the second Hopf bifurcation. Each of the two predicts the survival of three species, the ones, which are found inside the domains. Each of these fixed points is, of course, a single-site fixed point, so in principle a subset of the nodes of the grid can individually approach one of the two fixed points, while the complementary set of the nodes would approach the other fixed point. However, as a transient we see two well separated domains with either even or odd species. At the interfaces between them all six species are present and oscillate with small amplitude oscillations, caused by the first Hopf bifurcation of the six-species coexistence fixed-point, where it became a saddle. Which one of the domains wins the effective (2,1)-game in the end, where a single domain with all its three species survives, depends on the initial conditions and on the fact that diffusion is included; the mere stability analysis only suggests that six species at a site destabilize the interface between domains with either even or odd species.  In fact, the numerical integration and the Gillespie simulations both show that one domain gets extinct if the lattice size is small enough and/or the diffusion fast enough. As long as the two fixed points are stable, the (3,1)-game is played at each site of a domain in the sense of coexisting three species, which are not chasing each other, related to the neighboring sites only via diffusion, without forming any patterns. Patterns are only visible at the interface of the domains as a remnant of the unstable six-species coexistence fixed point.
\item  $\kappa/\gamma>2$: the third regime, which is of most interest for pattern formation.
Starting from random initial conditions, the species segregate first into two domains, each consisting of three species, one with species 1,3, and 5, the second one with species 2, 4, and 6, and inside both domains the three species play a rock-paper-scissors game, chasing each other, since the two fixed points of the $FP^3$ group became unstable at the second Hopf bifurcation. Due to the interactions according to an effective $(2,1)$-game at the interfaces of the domains (here with either two or four species coexisting), one of the domains will also here get extinct, including the involved three species, while the remaining three survive. Which domain survives  depends also here on the initial conditions. As we shall see, the temporal trajectories of the concentrations of the three species in the surviving domain show that they still explore  the vicinity of the second Hopf bifurcation from time to time, while they otherwise are attracted by the heteroclinic cycle. The three species in the surviving domain live the longer, the larger the grid size is, in which the species continue playing (3,1).  In contrast to the second regime, however, two of the three species in the surviving domain will get extinct as well, and a single one remains in the end. This extinction is caused by fluctuations in the finite population in the stochastic simulation or by the numerical integration on a spatial grid with finite numerical accuracy, respectively.
\end{itemize}
So the linear stability analysis indicates options for when we can expect oscillatory trajectories: it is  the Hopf bifurcations in the (6,3)-game for the $FP^2$ and $FP^3$ fixed points that induce the creation of limit cycles, which here lead to the {\it formation of spirals} in space in the third regime and only temporary patterns at the interfaces in the second regime, before the system converges to one of the  $FP^3$ fixed points.

Moreover, it is the two $FP^3$ fixed points in the (6,3)-game that correspond to the {\it formation of two domains}. In both the (3,2) and the (6,3)-games, one of these fixed points will be approached as a collective fixed point (shared by all sites of the grid), while the domain corresponding to the other one gets extinct, and patterns are seen if this fixed point is unstable. So in the (6,3)-game the existence of domains including their very composition is due to two stable (second regime) or unstable (third regime) fixed points. Their coexistence is in both regimes transient. In the second regime three species will survive in the end, because the three-species coexistence-fixed point is stable, and it would need a large fluctuation to kick it towards a 1-species unstable fixed point. In contrast, only one species will survive in the third regime, where the same fixed point is unstable. Obviously here it does not need a rare, large fluctuation to kick the system towards the 1-species unstable fixed point, as we always observed a single species to survive in the end, both in the Gillespie simulations and the numerical integration in a relatively short time.

We should mention, however, that from our Gillespie simulations we cannot exclude that after all, a large fluctuation  would also kick the system in the second regime from its metastable state towards one of the unstable 1-species fixed points  as well as in the first regime  to either one of the two three-species fixed points, or to one of the six 1-species fixed points, when the six-species fixed point is stable in the deterministic limit. So far we have not searched for these rare events, in which two, three or five species would get extinct, respectively.
\vskip5pt
\textbf{Numerical solutions of the (6,3)-game.}

\begin{figure}[tp]
	\begin{center}
		\includegraphics[scale=1]{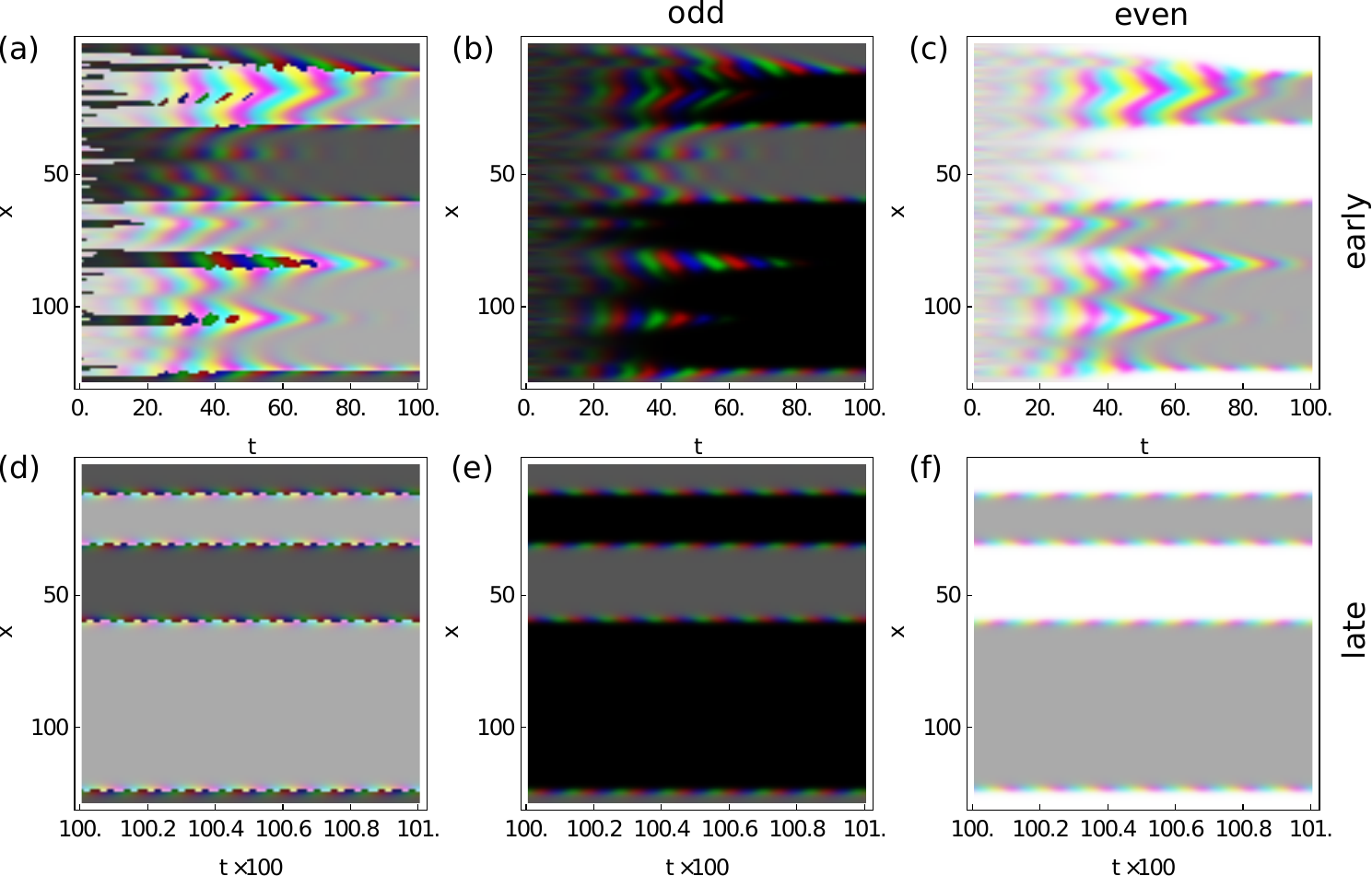}
	\end{center}
	\caption{Evolution of the (6,3)-game in the second regime in one dimension. The parameters are $ \gamma = \kappa = 1 $, $ \delta/dx = 0.1 $, and $ \kappa = 1.5 $. The right and middle columns show the species of each domain separately.
For further explanations see the text.}\label{(6,3)_array_DOM}
\end{figure}

In the following we show evolutions of species concentrations in space and time for parameters, chosen from the second and third regime of the (6,3)-game. These solutions are obtained from the numerical integration of Eq.~\ref{eq:MF(6,3)}.
For the representation on a lattice we will use the following procedure to visualize site occupation: Odd species are represented by the rgb-(red, green, blue) color scheme, while even species are represented by  cmy-colors (cyan, magenta, yellow). The three numbers of species $(r,g,b)$, or $(c,m,y)$, divided by the total sum of all species at the site, give a color in the rgb-, or cmy-spectrum that results from a weighted superposition of individual colors, where the weights (color intensities) depend only on the ratios of occupation numbers, rather than on absolute numbers. Moreover, we display the rgb-color scheme if odd species make up the majority at a site and the cmy-scheme otherwise. We should note that a well mixed occupation of  odd (even) species leads to a dark (light) gray color in these color schemes.
Figure~\ref{(6,3)_array_DOM} shows coexisting domains with oscillations at the interfaces in the second regime.  To justify the visualization of data according to the ``majority rule", we show even and odd species also separately in the two right panels. This way we can see the transitions at the interfaces  of the domains between even and odd species more clearly. The light (dark) gray domain corresponds to a well mixed occupancy with even (odd) species, respectively. On the boundaries of the domains all six species are present, and if we zoom into the boundary, we can see small amplitude oscillations caused by the Hopf bifurcation of the 6-species coexistence-fixed point, see figure~\ref{(6,3)_tx_DOM}. Figures~\ref{(6,3)_array_DOM} (a)-(c) show the evolution of the system at the first 100 t.u. at which time the domains are already starting to form. In panel (a) it is seen how the transient patterns, generated by transient small domains, shortly after the domains disappear also fade away, so that the transient patterns are generated by oscillations at the interfaces. The figure also reminds to the early time evolution of condensate formation in a zero-range process, where initially many small condensates form, which finally get absorbed in a condensate that is located at a single site with macroscopic occupation in the thermodynamic limit. Here initially many small and short-lived domains form, which get first absorbed into four domains as seen in the figure, but later end up in a single domain  with three surviving species. So we see a ``condensation" in species space, where three out of six species get macroscopically occupied as a result of the interaction, diffusion and an unstable interface, while the remaining three species get extinct,  so that the symmetry between the species in the cyclic interactions with identical rates gets dynamically broken.
\begin{figure}[tp]
	\begin{center}
		\includegraphics[scale=1]{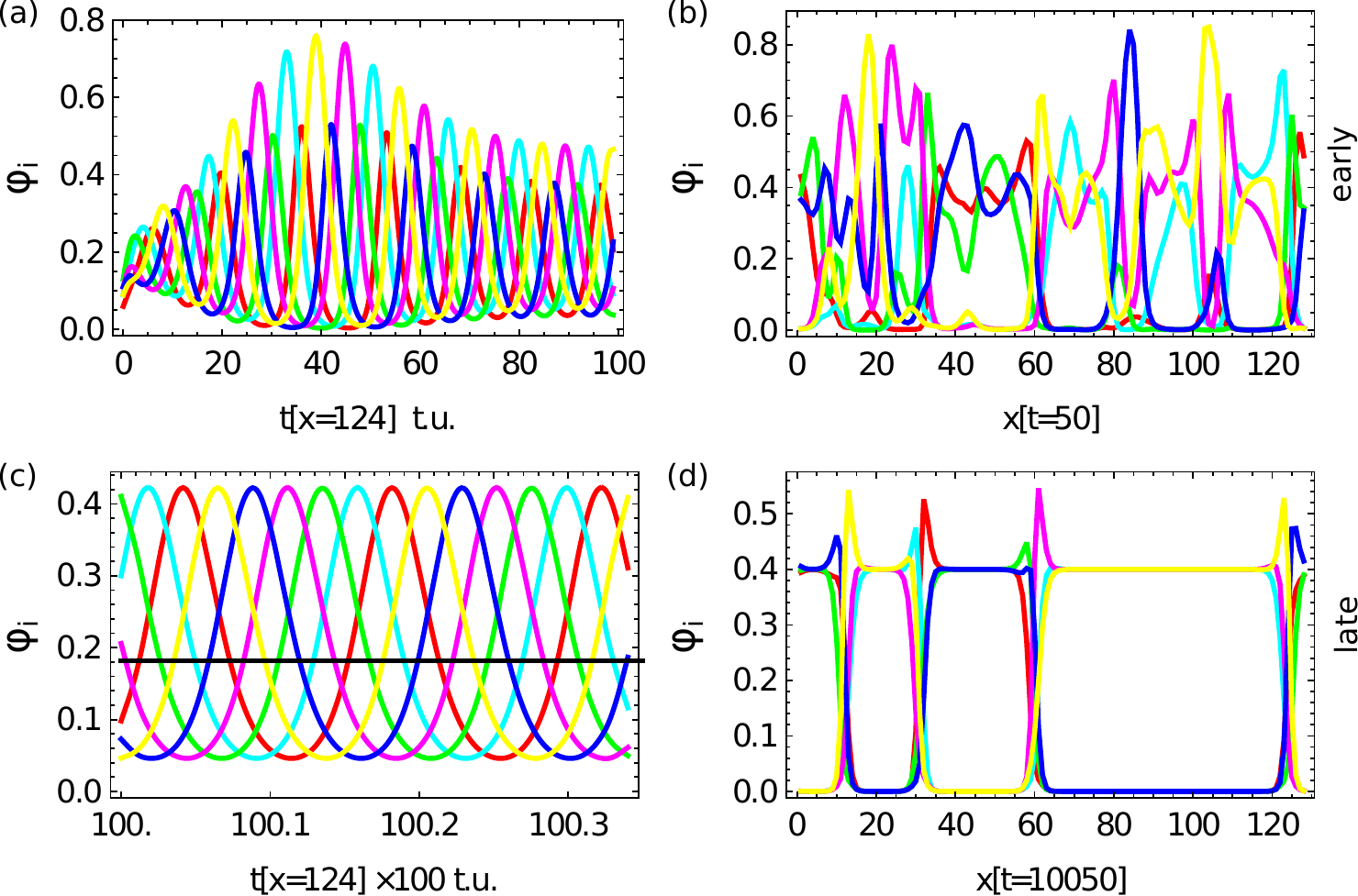}\\
	\end{center}
	\caption{Trajectories of all six species corresponding to figure~\ref{(6,3)_array_DOM} at an interface. Red (1), green (3), and blue (5) represent odd species, cyan (2), magenta (4), and yellow (6) even species, from smaller to larger labels, respectively. (a) and (b) show temporal and spatial trajectories, respectively, at the beginning of the integration, corresponding to  (a)-(c) in figure~\ref{(6,3)_array_DOM}, while (c) and (d) refer to late times. For further explanations see the text.}\label{(6,3)_tx_DOM}
\end{figure}

Panels (d)-(f) show the evolution from 10000-10100 time units (t.u.). The displayed domains were checked to coexist numerically stable up to $10^6$ t.u., while for smaller lattices and faster diffusion one domain gets extinct. Figures~\ref{(6,3)_array_DOM} (a) and (d) should be compared with figures~\ref{(6,3)1D_2} (a) and (b) of the Gillespie simulations, respectively.

Figure~\ref{(6,3)_tx_DOM} shows the corresponding oscillating concentration trajectories at early (a) and late (c) times at a site of an interface (x=124), where all six species oscillate around the coexistence-saddle fixed point, as indicated by the horizontal black line in (c), while the spatial dependence at (b) (early) and (d) (late) times displays the domain formation due to two stable fixed points, corresponding to figures~\ref{(6,3)_array_DOM} (a) and (d), respectively, so that the oscillations are restricted to the interfaces.
\begin{figure}[tp]
	\begin{center}
		\includegraphics[scale=1]{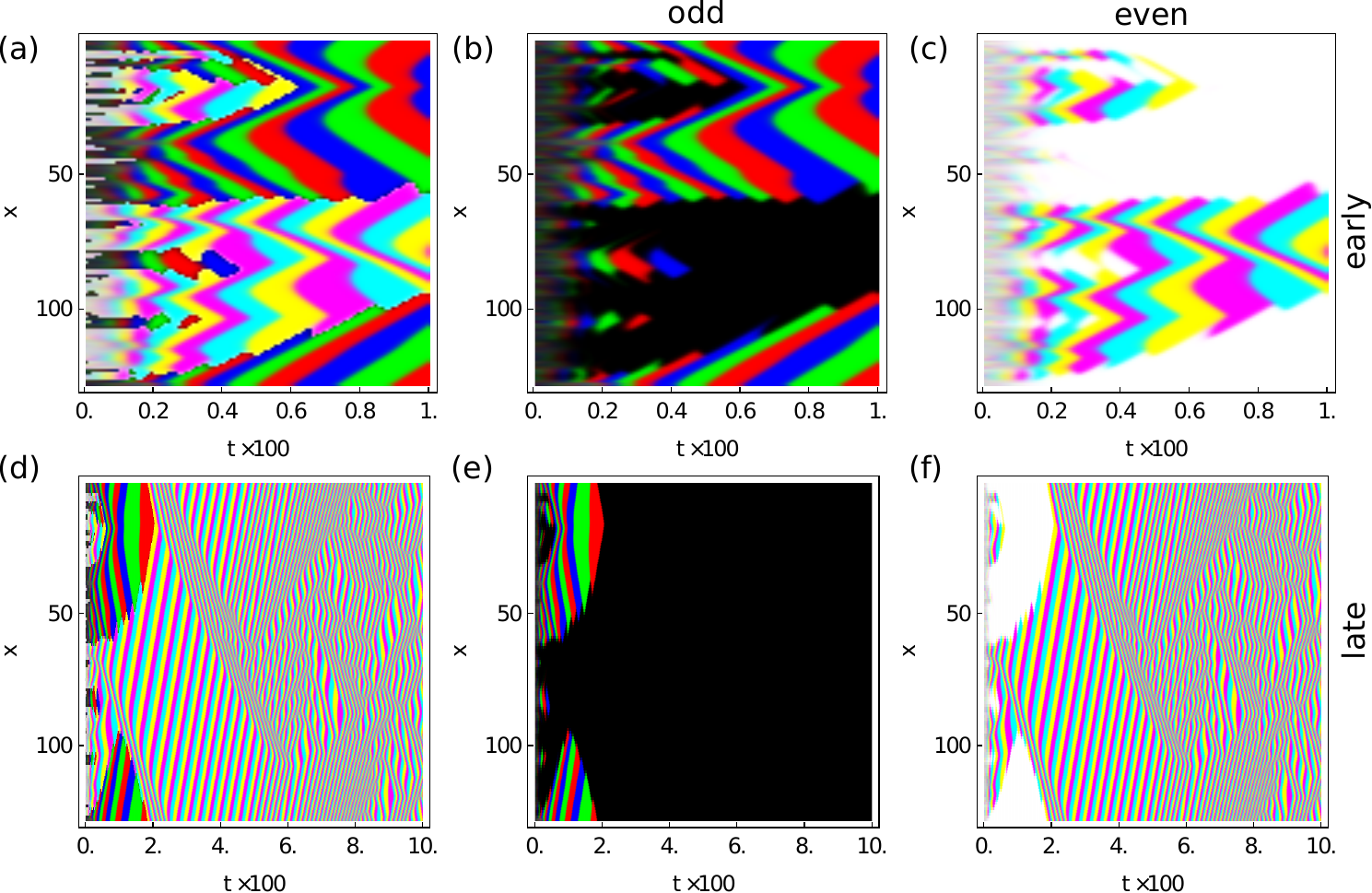}
	\end{center}
	\caption{Evolution of the (6,3)-game in the third regime in one-dimension. The parameters are $ \gamma = \kappa = 1 $, $ \delta/dx = 0.1 $, and $ \kappa = 4.0 $. The middle and right columns show the species of each domain separately. For further explanations see the text. }\label{(6,3)_array_OSC}
\end{figure}

The evolution of the (6,3)-game in the third, oscillatory regime in one dimension is shown in figure~\ref{(6,3)_array_OSC}. Species are represented in the same way as in figure~\ref{(6,3)_array_DOM}. Panel (a)-(c) show the evolution of the system in the first 100 t.u.,  (d)-(f) in the first 10000 t.u. The two stable fixed points from the second regime became unstable (saddles) through the second Hopf bifurcation. As in the second regime, at the beginning of the integration there is a separation of odd and even species, but at the same time they start to chase  each other, resulting in oscillatory behavior in space and time. Here we see no longer traces of the limit cycle around the six-species coexistence fixed point as in the second regime, since no sites have six species coexisting, even not for a short period of time. At the interfaces between even and odd species usually three species coexist, either two odd and one even, or vice versa, two even and one odd, but these mixtures are not stable, as these 3-species coexistence-fixed points in the deterministic limit are saddles. It also happens that just two or four species coexist at the interface, but also their coexistence-fixed points are saddles. Therefore also here the coexistence of domains is not stable,  only one of them survives, and which one depends on the initial conditions, resulting in the extinction of three either odd or even species. In view of Gillespie simulations, figures~\ref{(6,3)_array_OSC} (a) (early times) and (b) (late times) should be compared with figures~\ref{(6,3)1D_1} (a) (early) and (b) (late), respectively.

Figure~\ref{(6,3)_tx_OSC} (a) shows the evolution in time at late times, when only one domain survives.  All three species oscillate between zero and one, corresponding to the heteroclinic cycle. From time to time the trajectories are also attracted by the saddle-limit cycle, which is created by the second Hopf bifurcation of the three species-fixed point (black line) as indicated by the small amplitude oscillations. Apart from the amplitude, the heteroclinic and saddle-limit cycles differ  in their frequency: the saddle-limit cycle has a higher frequency than the heteroclinic cycle. Panel (b) shows the spatial trajectories at the beginning of the integration when both domains still coexist. Yet we see no mixing of all six species at a single site, the 6-species coexistence-fixed point is no longer felt in this regime.

\begin{figure}[tp]
	\begin{center}
		\includegraphics[scale=1]{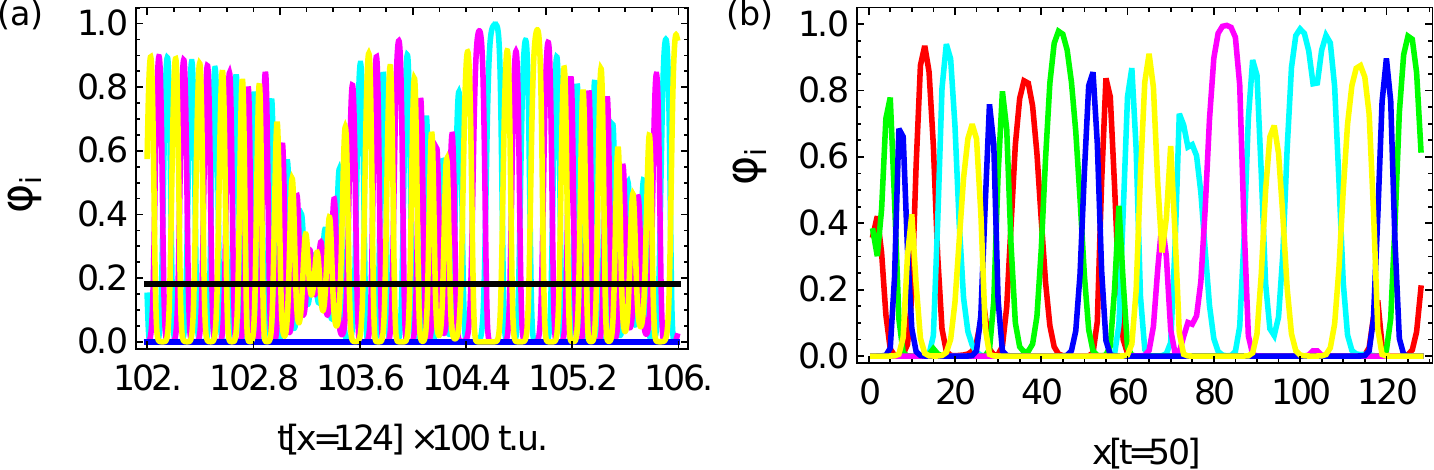}\\
	\end{center}
	\caption{In correspondence to figure~\ref{(6,3)_array_OSC} (a) temporal trajectories of the surviving domain at late times in the interval 10200-10600 t.u. when only even species exist, and (b) spatial trajectories  at early times when still both domains exist. For further explanations see the text.}\label{(6,3)_tx_OSC}
\end{figure}

As we see from figure~\ref{(6,3)_array_OSCend}, the numerical integration evolves to one of the saddles  after having spent a finite time on the heteroclinic cycle and not according to the analytical prediction, where it were only in the infinite-time limit that the trajectory would get stuck in one of the saddles, which are connected by the heteroclinic cycle. According to  figure~\ref{(6,3)_array_OSCend}(a) all trajectories get absorbed in one (the pink one) of the 1-species saddles already at finite time as a result of the finite accuracy of the numerical integration. Yet figure~\ref{(6,3)_array_OSCend} (b) shows the characteristics of a heteroclinic cycle at finite time: The dwell time of the trajectory in the vicinity of the 1-species saddles gets longer and longer in each cycle, before it fast moves towards the next saddle in the cycle. This escape stops after a finite number of cycles, when the concentration of two of the three  species are zero within the numerical accuracy, and therefore no ``resurrection" is possible.

\begin{figure}[htbp]
	\begin{center}
		\includegraphics[scale=1]{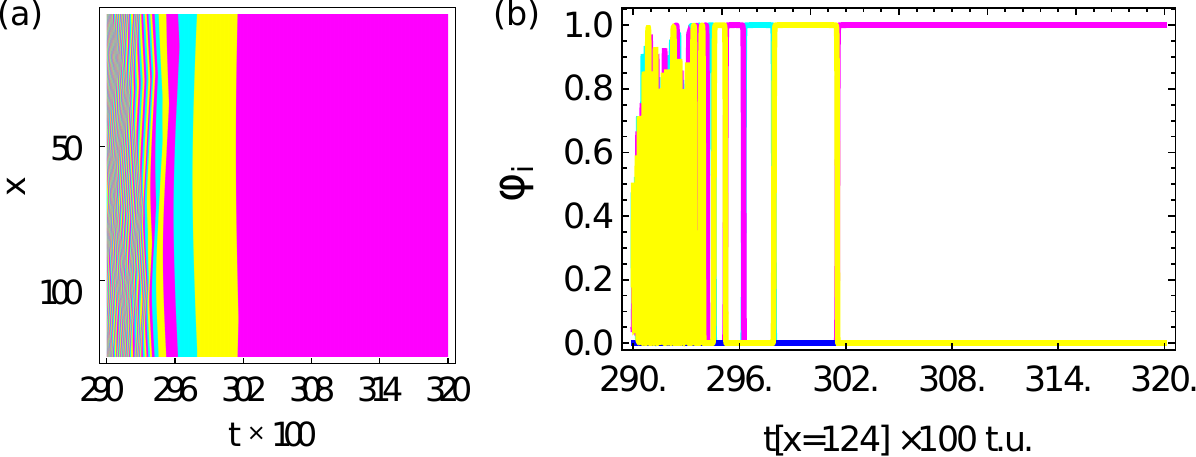}
	\end{center}
	\caption{Extinction of all but one species upon approaching the heteroclinic cycle, (a) (x,t)-diagram, (b) species concentrations as a function of time. For further explanations see the text.}\label{(6,3)_array_OSCend}
\end{figure}

%%%%%%%%%%%%%%%%%%%%%%%%%%%%%%%%%%%%%%%%%%%%%%%%%%%%%%%%%%%%%%%%%%%%%%%%%%%%%%%%%%%%%%%%%%%%%%%%%%%

\section{Numerical Methods and Results}\label{sec_numerical}
Going back to the set of reactions, in this section we describe their Gillespie simulations.
We solve the system (\ref{eq:rec_sys})
by stochastic simulations on a regular square lattice as well as on a one-dimensional ring, using the Gillespie algorithm~\cite{gillespie}, combined with the so-called next-subvolume method~\cite{elf}. This method is one option to generalize Gillespie simulations to spatial grids. We choose periodic boundary conditions on a square $L\times L$ lattice or on a ring with $L$ nodes. In our case nodes, or synonymously sites, represent subvolumes. All reactions except of the diffusion happen between individuals on the same site (in the same subvolume), and a diffusion reaction is a jump of one individual to a neighboring site. One event can change the state of the system of only one (if a reaction happens) or two neighboring (if  a diffusion event happens) subvolumes. At each site the initial number of individuals of each species is chosen from a Poisson distribution $P \left( \{n\} ;0 \right)=\underset{\alpha,i}{\prod}\left( \frac{\overline{n}^{n_{\alpha,i}}_{\alpha,0}}{n_{\alpha,i}!} e^{-\overline{n}_{\alpha,0}} \right)$, with a mean $\overline{n}_{\alpha,0}$, which is randomly chosen for each species $\alpha$.

In the next-subvolume method we assign the random times of the Gillespie algorithm to subvolumes rather than to a specific reaction. To each subvolume, or site, we assign  a time $\tau$, at which one of the possible events, in our case reactions (predation, birth or death), or diffusion, will happen. The time $\tau$ is calculated as $\tau = -\ln(rn)/r_{total}$, where $rn$ is a random number generated from a uniform distribution between 0 and 1. The total rate $r_{total}$ depends on the reaction rates and the number of individuals which participate in the event. Events happen at sites in the order of the assigned times $\tau$. Once it is known at which subvolume the next event happens, the event (reaction or diffusion) is chosen randomly according to the specified reaction rates.

We start the simulations with initial conditions from a Poisson distribution such that each site of the entire lattice is well mixed with all species. We want to study the dynamics of the system in a parameter regime, where we expect pattern formation. From the linear stability analysis of the mean-field system we expect stable patterns in the regime without stable fixed points in the (6,3)-game, as long as three species are alive, and transient patterns in the (6,3)-game for  coexisting stable fixed points.
We use the same color scheme as we used before to visualize the numerical solutions of the mean-field equations.
As one Gillespie step (GS) we count one integration step here.

Our results confirm the predictions from the mean-field analysis: there are on-site oscillations in time in a limit-cycle regime. There are also oscillations in space, which form spirals on two-dimensional lattices. If the evolution approaches the stochastic counterpart of a stable fixed point in the deterministic limit, we call it shortly  ``noisy fixed point",  where the trajectories fluctuate around a value that is the mean field-fixed point value multiplied by the parameter $V$ as defined before. Of particular interest is the influence of the diffusion in relation to the ratio $\kappa/\gamma$ on the patterns. It was the ratio of $\kappa/\gamma$ that determines the stability of the fixed points. As mentioned earlier, the value of $\delta k^2$, which enters the stability analysis, can extend the stability regime. So in the Gillespie simulations it is intrinsically hard to disentangle the following two reasons for the absence of patterns in the case of fast diffusion: either the stability regime of a fixed point with only one surviving species is extended, or the diffusion is so fast, that the extension of visible patterns is larger than the system size, so that a uniform color may just reflect the homogeneous part within a large pattern.
All the mean-field-fixed points are proportional to the value of the parameter $\rho$. If this value is much larger than $\kappa$ and $\gamma$, the fixed-point value is very large. This leads to a large occupation on the sites, which slows down the formation of patterns. The reason is that the number of reactions, which are needed for the system to evolve to stable trajectories, either to oscillations, or to fixed points, increases with the number of individuals in the system.

\vskip5pt

We study the stochastic dynamics of a (6,3)-game in regimes, for which we expect pattern formation, i.e. for $\gamma<\kappa$. When the coexistence-fixed point $FP^2$ becomes unstable at $\gamma=\kappa$, we find the formation of two domains, each consisting of three species, one domain containing odd species, in the figures represented by shared colors red, green, and blue in the rgb-color scheme. The other domain consists of even species, represented by shared colors of cyan, magenta, and yellow in the cmy-color representation, see figure~\ref{(6,3)2D}. Inside the domains the three species play the (3,1)-game and form spiral patterns. We have checked that the domains in figure~\ref{(6,3)2D} are not an artefact of the visualization, and determined, for example, the occupancy on a middle column of the lattice (not displayed here).
On sites with oscillations of species from one domain, there is a very small or no occupation of species of the second domain, confirming the very existence of the domains.
The time evolution of the six species on two sites, chosen, for example, from the middle column of the lattice confirm that the species' trajectories  oscillate in time, reflecting the stable limit cycles in the deterministic limit.

Here a remark is in order as to whether radii, propagation velocity or other features of the observed spiral patterns can be predicted analytically. While  spiral patterns in spatial rock-paper-scissors games were very well predicted via a multi-scale expansion in the work of \cite{mobilia1,mobilia2}, we performed a multi-scale expansion (see, for example \cite{bookkuramoto}) to derive amplitude equations for the time evolution of deviations from the two unstable fixed points, which lose their stability at the two Hopf bifurcations. However, the resulting amplitude equations differ from Ginzburg-Landau equations by a missing imaginary part, which can be traced back to the absence of an explicit constraint to the occupation numbers on sites and the absence of a conserved total number of individuals. As a result, the amplitude equations only predict the transient evolution as long as the trajectory is in the very vicinity of the unstable fixed point, but cannot capture the long-time behavior, which here is determined by an attraction towards the heteroclinic cycle that is responsible for the spiral patterns in our case. So it seems to be this non-local feature in phase space that the multi-scale expansion about the Hopf bifurcation misses.

For a further discussion of how the patterns depend on the choice of parameters  we shall focus on the results on a one-dimensional lattice, since the simulation times are much longer for two dimensions. (In two dimensions, the period of oscillations is as long as about one fifth of the $2^{30}$ Gillespie steps.)

\begin{figure}[tp]
	\begin{center}
		\includegraphics[scale=0.9]{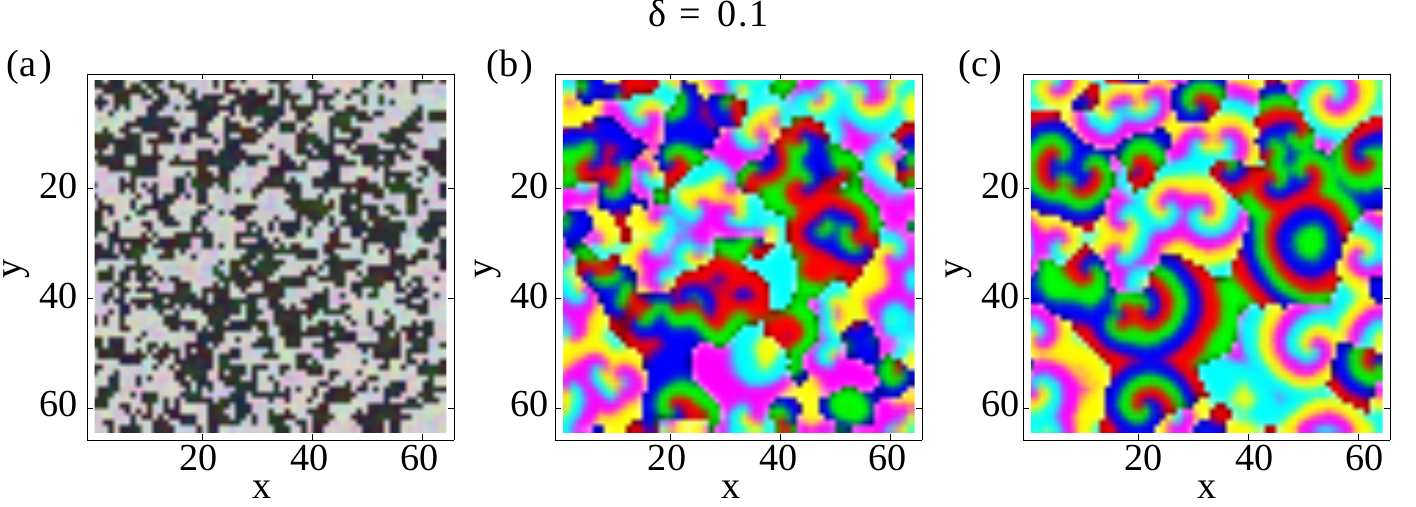}
		\end{center}
	\caption{Pattern formation for a (6,3)-game on a two-dimensional $(64\times 64)$-lattice for weak diffusion and far from the bifurcation point. Snap shots are taken at  $1000\cdot2^{15}$ (a), $10000\cdot2^{15}$ (b), and $32000\cdot2^{15}$ (c) GS. Two domains are formed, each containing three species, indicated by the different color groups. These species play a (3,1)-game inside the domains and evolve spiral patterns. The parameters are $\rho=1$, $\kappa=1$, $\delta=1$, and $\gamma=0.2$.
}\label{(6,3)2D}
\end{figure}

\begin{figure}[tp]
	\begin{center}
		\includegraphics[scale=1]{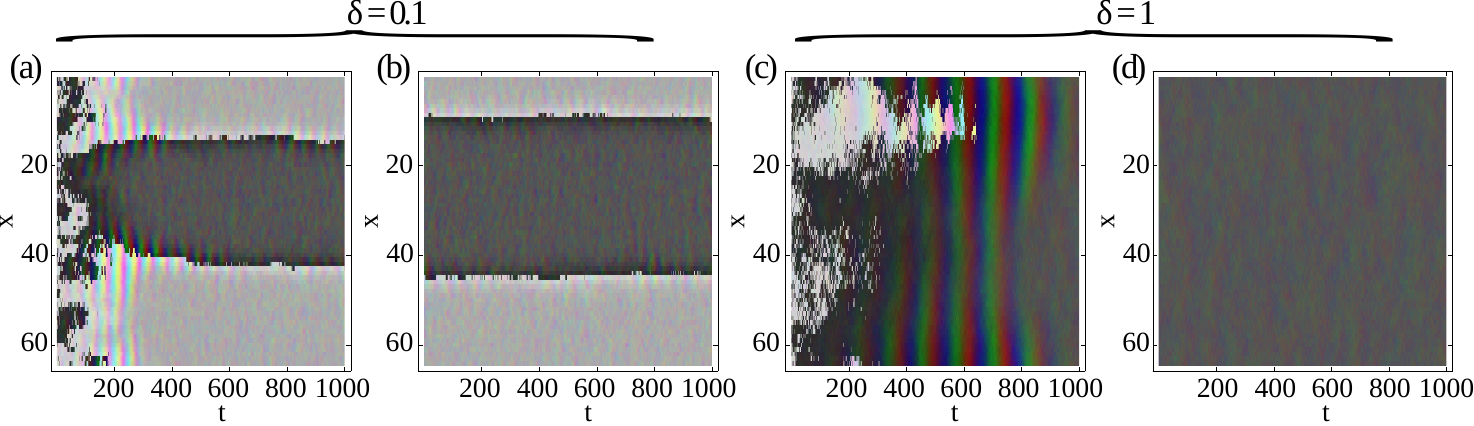}
		\end{center}
	\caption{Pattern formation in the (6,3)-game on a one-dimensional lattice of 64 sites for $\kappa/2<\gamma<\kappa$, that is in the second regime, where the coexistence-fixed points $FP^3$ are stable, for weak diffusion $\delta=0.1$ ((a) and (b)), and strong diffusion $\delta=1$ ((c) and (d)). For both strengths of the diffusion domains form. In the case of weak diffusion no extinction of domains is observed within the simulation time of $2^{30}$ GS. For strong diffusion, one domain goes extinct after $800\cdot2^{15}$ GS. Initially,  oscillatory patterns appear as remnants of many interfaces between small domains, where within the interfaces six species oscillate due to the unstable 6-species coexistence-fixed point, which fade away with time. This confirms the analytical results that in this parameter regime the $FP^3$-fixed points with $2\times 3$ coexisting species  are both stable, leading to the black color in (c) and (d) for the one surviving domain. Panel (a) shows the time evolution on a lattice for the time interval $(0-1000)\cdot2^{15}$, (b) for $(30000-31000)\cdot2^{15}$, (c) for $(0-1000)\cdot2^{15}$, and (d) for $(10000-11000)\cdot2^{15}$ GS. The parameters are $\gamma=\rho=0.5$, $\kappa=0.6$.}\label{(6,3)1D_2}
\end{figure}

% one-dimesional lattice
\begin{figure}[htbp]
	\begin{center}
		\includegraphics[scale=1]{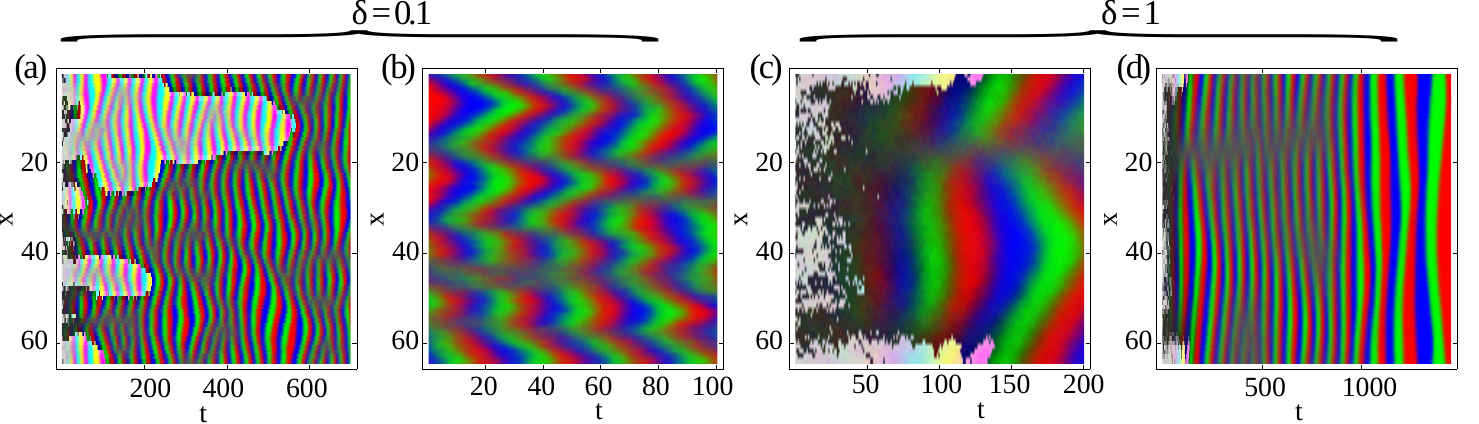}
		\end{center}
	\caption{Pattern formation in the (6,3)-game on a one-dimensional lattice of size 64, shown  on a space-time grid, for $\gamma<\kappa/2$, that is, in the third regime, where the coexistence-fixed points $FP^3$ are unstable, weak diffusion $\delta=0.1$ ((a) and (b)), and strong diffusion $\delta=0.1$ ((c) and (d)). The parameters are $\gamma=\rho=0.5$ and $\kappa=1.2$. Two domains form, of which one goes extinct after $600\cdot2^{15}$ GS in the case of weak diffusion and $150\cdot2^{15}$ GS in the case of strong diffusion. The surviving domain keeps playing the (3,1)-game. For weak diffusion no further extinction is observed for the simulation time of $2^{30}$ GS, while for strong diffusion, an extinction of all but one species, here the red one, happens after $1470\cdot2^{15}$ GS. Panel (a) shows patterns in a time interval of $(0-700)\cdot2^{15}$, (b) for $(30000-30100)\cdot2^{15}$, (c) for $(0-200)\cdot2^{15}$, and (d) for $(0-1470)\cdot2^{15}$ GS.}\label{(6,3)1D_1}
\end{figure}

As to diffusion, for stronger diffusion patterns are more homogeneous, extinction events happen faster,  sometimes they happen only for sufficiently strong diffusion, see figure~ \ref{(6,3)1D_2}. The extinction time depends also on the $\kappa/\gamma$-ratio, i.e. if the ratio is from the interval (i) $1<\kappa/\gamma<2$, where $FP^3$ is stable and $FP^2$ is unstable, or, (ii) from $\kappa/\gamma>2$, where both fixed points are unstable.

For case (i), the $FP^3$-fixed points are stable, yet at the beginning of the simulations the dynamics shows oscillatory behavior, caused by the interfaces between the small domains, where six species feel the unstable coexistence-fixed point at $\gamma=\kappa$; but after about $500\cdot2^{15}$ Gillespie steps for weak diffusion, and $1000\cdot2^{15}$ Gillespie steps for strong diffusion, and for our choice of parameters, the patterns fade away and the system evolves to a homogeneous state in both domains as long as they coexist, see figures~\ref{(6,3)1D_2} (b) and \ref{(6,3)1D_2} (d). The closer the system is to the bifurcation point $\gamma=\kappa$, the longer live the oscillatory patterns, the stronger feels the system the unstable 6-species fixed point.

Figure~\ref{(6,3)1D_2} (a) should be compared with the corresponding mean-field solution of figure~\ref{(6,3)_array_DOM} (a) at early times and figure~\ref{(6,3)1D_2} (b) with figure~\ref{(6,3)_array_DOM} (d) at late times, from which we see that the mean-field solutions reproduce the qualitative features including the transient patterns.

In case (ii), the third regime, domains get faster extinct, both domains play a (3,1)-game inside the domains. After one domain gets extinct, the surviving one keeps on playing the (3,1)-game, until only one species survives. Here one should compare figure~\ref{(6,3)1D_1} (a) with the corresponding mean-field solution of figure~\ref{(6,3)_array_OSC} (a) at early times and figure~\ref{(6,3)1D_1} (b) with figure~\ref{(6,3)_array_OSC} (d) at later times, where one is left with one domain and three species chasing each other; the final extinction of two further species is not visible in this figure due to the weak diffusion and the larger extinction time.

%%%%%%%%%%%%%%%%%%%%%%%%%%%%%%%%%%%%%%%%%%%%%%%%%%%%%%%%%%%%%%%%%%%%%%%%%%%%%%%%%%%%%%%%%%%%%%%

\section{Conclusions and Outlook}\label{sec_conclusions}
Beyond the emerging dynamically generated time-and spatial scales, the most interesting feature of the (6,3)-game is the fact that the rules of the game, specified initially as (6,3), dynamically change to effectively (2,1) and (3,1) as a result of spatial segregation. In view of evolution, here the rules of the game change while being played. They change as a function of the state of the system if the state corresponds to the spatial distribution of coexisting species over the grid.

In preliminary studies, we investigated the $(27,17)$-game with the following set of coexisting  games in a transient period: From a random start we observe segregation towards nine domains playing $(9,7)$ with each other and inside the domains again the $(3,1)$ game. From the superficial visualization of Gillespie simulations, this system looks like a fluid with whirling ``vortices", where the (3,1)-game is played inside the domains. We expect a rich variety of games with new, emerging, effective rules on a coarser scale, if we not only increase the number $N$ of species, or release the restriction to cyclic predation, but allow for different time scales, defined via the interaction rates. So far we chose the same rates for all interactions and always started from random initial conditions.

We performed a detailed linear stability analysis, which together with the numerical integration of the mean-field equations reproduced all qualitative features of the Gillespie simulations, even extinction events. That the mean-field analysis worked so well is due to the ultralocal implementation of the interactions, so that the spatial dependence enters only via diffusion. The  stability analysis revealed already a rather rich structure with 12 groups of in total 64 fixed points for the (6,3)-game. We focussed on coexistence-fixed points of six, three or one species.

Along with the fixed points' repulsion or attraction properties we observed three types of extinction, whose microscopic realization is different and deserves further studies:

(i) In the second regime of the (6,3)-game, both domains with either even or odd species are in principle stable, as long as they are not forced to coexist. We have seen a spatial segregation towards a domain with only even and one with only odd species, occupying the sites. At the interface between both domains, six species cannot escape from playing the (6,3)-game. Since the 6-species coexistence-fixed point is unstable, the unstable interface seems to be the driving force to initiate the extinction of one of the two domains including its three species, since interface areas should be reduced to a minimal size. From the coarse-grained perspective, one domain preys on the other domain, which is a (2,1)-game.

(ii) In the third regime of the (6,3)-game, the domain structure in odd and even domains is kept, but in the interior of the domains the species follow heteroclinic cycles, which explain the patterns of three species, chasing each other, inside each domain.
At the interface between the domains, two to four species coexist at a site, but for small enough diffusion, coexistence-fixed points of the respective species are always saddles, so also here the instability of the interfaces seems to induce their avoidance, leading
again to the extinction of one of the two domains. So from the coarse-grained perspective, again a (2,1)-game is played between the domains.\\
It should be noticed that in contrast to systems, where the fate of interfaces between domains is explained in terms of the competition between free energy and interface tension, here the growth of domains and the reduction of interfaces are traced back to the linear stability analysis of the system in the deterministic limit, which is conclusive for the dynamics.

(iii) The third type of extinction event was the extinction of two species, when the individual trajectories move either in the vicinity of, or  along a heteroclinic cycle, and either a fluctuation from the Gillespie simulations, or the finite numerical accuracy on the grid (used for integration) captured the trajectory in one of the 1-species saddles.

We have not studied rare large fluctuations, which could induce other extinction events and kick the system out of the basin of attraction from the 6- or 3-species stable coexistence-fixed points when stochastic fluctuations are included. Neither have we measured any scaling of the extinction times with the system size or of the domain growth with the system size. This is left for future work.\\
Furthermore, for future work it would be challenging to derive and predict the domain formation on the coarse scale from the underlying $(6,3)$-game on the basic lattice scale in the spirit of the renormalization group, here, however, applied to differential equations rather than to an action.

\section{Acknowledgments}
One of us (D.L.) is grateful to the German Research Foundation (DFG)(ME-1332/25-1) for financial support during this project. We are also indebted to the German Exchange Foundation (DAAD)(ID 57129624)for financial support of our visit at Virginia Tech Blacksburg University, where we would like to thank Michel Pleimling for valuable discussions. We are also indebted to Michael Zaks (Potsdam University) for useful discussions.\\

%%%%%%%%%% Merge with supplemental materials %%%%%%%%%%
\pagebreak
\widetext
\begin{center}
\textbf{\large Supplemental Material: Rock-paper-scissors played within competing domains in predator-prey games}
\end{center}
%%%%%%%%%% Merge with supplemental materials %%%%%%%%%%
%%%%%%%%%% Prefix a "S" to all equations, figures, tables and reset the counter %%%%%%%%%%
\setcounter{equation}{0}
\setcounter{figure}{0}
\setcounter{table}{0}
\setcounter{page}{1}
\makeatletter
\renewcommand{\theequation}{S\arabic{equation}}
\renewcommand{\thefigure}{S\arabic{figure}}
\renewcommand{\bibnumfmt}[1]{[S#1]}
\renewcommand{\citenumfont}[1]{S#1}
%%%%%%%%%% Prefix a "S" to all equations, figures, tables and reset the counter %%%%%%%%%%
\setcounter{section}{0}
\vspace{1cm}
We add a detailed bifurcation analysis of the (3,1)- and (3,2)-games with only spiral ((3,1)) or only domain ((3,2)) formation, including the numerical integration of the mean-field solutions and the results of the Gillespie simulations.

\section{Stability analysis and numerical integration for the (3,1)-game}
\subsection{Stability analysis for the (3,1)-game}
The mean field equations for the (3,1)-game with homogeneous parameters are given by
\begin{eqnarray}\label{eq:MF(3,1)}
\frac{\partial\varphi_1}{\partial t} & = & \nabla^2\varphi_1 + \rho\varphi_1 - \gamma\varphi_1^2 - \kappa\varphi_3\varphi_1 \nonumber \\
\frac{\partial\varphi_2}{\partial t} & = & \nabla^2\varphi_2 + \rho\varphi_2 - \gamma\varphi_2^2 - \kappa\varphi_1\varphi_2 \nonumber \\
\frac{\partial\varphi_3}{\partial t} & = & \nabla^2\varphi_3 + \rho\varphi_3 - \gamma\varphi_3^2 - \kappa\varphi_2\varphi_3  \\
\end{eqnarray}
with Jacobian $J$ of the system~(\ref{eq:MF(3,1)})
\begin{equation}
	J=
	\begin{pmatrix}
		\rho-2\gamma\varphi_1^*-\kappa\varphi_3^* & 0 & -\kappa\varphi_1^*\\
		-\kappa\varphi_2^* & \rho-2\gamma\varphi_2^*-\kappa\varphi_1^* & 0 \\
		0 & -\kappa\varphi_3^* & \rho-2\gamma\varphi_3^*-\kappa\varphi_2^*
	\end{pmatrix}.
\end{equation}
The following table gives the list of all fixed points and the corresponding eigenvalues for the system~(\ref{eq:MF(3,1)}) without and with the spatial component.

\vspace{5mm}
{\footnotesize
\centering
\begin{tabular}{c|c|c|c}
	&	Fixed points & Eigenvalues for $D_\alpha=0$&  Eigenvalues for $D_\alpha\neq 0$ \\ \hline	
	$FP_1$ & 	(0, 0, 0) & ($\rho$, $\rho$, $\rho$) & ($\rho-\delta k^2$, $\rho-\delta k^2$, $\rho-\delta k^2$)\\ \hline
	$FP_2$ & ($\frac{\rho}{\gamma+\kappa}$, $\frac{\rho}{\gamma+\kappa}$, $\frac{\rho}{\gamma+\kappa}$) & ($-\rho \frac{\gamma+\kappa}{\gamma+\kappa}$, $-\rho \frac{2\gamma-\kappa\pm i\sqrt{3}\kappa}{2(\gamma+\kappa)}$) &  ($-\rho-\delta k^2$, $-\frac{\rho(2\gamma-\kappa)\pm i\sqrt{3}\kappa\rho}{2(\gamma+\kappa)}-\delta k^2$)\\ \hline
	$FP_3$ & ($\frac{\rho}{\gamma}$, 0, 0) & &\\
	$FP_4$ & (0, $\frac{\rho}{\gamma}$, 0) & ($-\rho$, $\rho$, $\rho\frac{\gamma-\kappa}{\gamma}$)& ($-\rho-\delta k^2$, $\rho-\delta k^2$, $\frac{\rho(\gamma-\kappa)}{\gamma}-\delta k^2$)\\
	$FP_5$ & (0, 0, $\frac{\rho}{\gamma}$) & &  \\\hline
	$FP_6$ & ($\frac{\rho}{\gamma}$, $\frac{(\gamma-\kappa)\rho}{\gamma^2}$, 0) & & \\
	$FP_7$ & ($\frac{(\gamma-\kappa)\rho}{\gamma^2}$, 0, $\frac{\rho}{\gamma}$) & ($-\rho$, $-\rho\frac{\gamma-\kappa}{\gamma}$, $\rho\frac{\gamma^2-\gamma\kappa+\kappa^2}{\gamma^2}$) & ($-\rho-\delta k^2$, $-\frac{\rho(\gamma-\kappa)}{\gamma}-\delta k^2$, $\frac{\rho(\gamma^2-\gamma\kappa+\kappa^2)}{\gamma ^2}-\delta k^2$) \\
	$FP_8$ & (0, $\frac{\rho}{\gamma}$, $\frac{(\gamma-\kappa)\rho}{\gamma^2}$) & & \\
\end{tabular}}

\vspace{5mm}

Let us next analyze the stability of the fixed points, first without the spatial component. The zero-fixed point $FP_1$ has all three eigenvalues positive for any choice of the positive parameters (we only consider positive parameters for this set of reactions), so it is always unstable in all directions. The second fixed point, the coexistence-fixed point $FP_2$, has one real eigenvalue that is always negative, so it corresponds to a stable direction, and two complex eigenvalues whose real parts change sign at $\kappa/\gamma=2$ through a supercritical Hopf bifurcation. For $\kappa/\gamma<2$ the fixed point is stable, otherwise it is unstable in the directions corresponding to the pair of complex conjugate eigenvalues.\\
The other fixed points are always saddles, but the number of stable/unstable directions depends on the ratio of $\kappa/\gamma$ and changes at $\kappa/\gamma=1$ through a transcritical bifurcation. Fixed points $FP_3$-$FP_5$, for which only one species is different from zero have always one stable ($-\rho$ eigenvalue) and one unstable ($\rho$ eigenvalue) direction. The third direction is stable for $\kappa/\gamma>1$, otherwise it is unstable. Three fixed points $FP_6$-$FP_8$, which correspond to the survival of two species, also have always one stable ($-\rho$ eigenvalue) and one unstable ($\rho\frac{\gamma^2-\gamma\kappa+\kappa^2}{\gamma^2}$ eigenvalue) direction. The third direction changes sign at the same point in phase space as in case of the previous fixed points, at $\kappa/\gamma=1$. The direction corresponding to this eigenvalue is stable for $\kappa/\gamma<1$, in contrast to the previous case. For example, for $\kappa/\gamma>1$, the fixed point $FP_3=(\frac{\rho}{\gamma}, 0, 0)$ has two stable and one unstable direction, while the fixed point $FP_6=(\frac{\rho}{\gamma}, \frac{(\gamma-\kappa)\rho}{\gamma^2}, 0)$ has one stable and two unstable directions. At $\kappa/\gamma=1$ the fixed points $FP_3$ and $FP_6$ collide and exchange the stability of one of the directions, so that for $\kappa/\gamma<1$ $FP_3$ has one stable and two unstable directions, while $FP_6$ has two stable and one unstable direction. The same scenario happens for other fixed points of the same type, $FP_4$ collides with $FP_8$ and $FP_5$ with $FP_7$.\\
After the collision, the saddles $FP_3-FP_5$ are connected by a heteroclinic cycle.  For $\kappa/\gamma<2$ the heteroclinic cycle is  repelling, the system is permanent\footnote{According to ~\cite{aubin}, the system, described by a set of dynamical equations, is called permanent, if the boundary is an unreachable ``repellor", or equivalently, if there exists a compact subset in the interior of state space, where all orbits starting from the interior eventually end up.}. In our system this  means that the heteroclinic cycle repels all trajectories towards the stable coexistence-fixed point and bounds the phase space towards this interior. All trajectories evolve to this fixed point, regardless of the initial conditions. For $\kappa/\gamma>2$, the heteroclinic cycle becomes an attractor, and for all initial conditions, except for those that lie on a diagonal of the phase space (connecting the zero-species fixed point with the coexistence-fixed point), the system evolves to the heteroclinic cycle. Both results (repelling property for $\kappa/\gamma<2$ and attracting for $\kappa/\gamma>2$ of the heteroclinic cycle) follow from the proof in \cite{hofbauer}. \\
The numerical integration shows the typical behavior for a heteroclinic cycle, the trajectory oscillates between the three saddles, staying longer and longer in their vicinity, as time passes by, until  it gets absorbed  by one of the saddles due to finite numerical accuracy.\\ Figure~\ref{(3,1)traj} shows the stability of the fixed points and typical trajectories, when the initial conditions are close to the zero-fixed point $FP_1$.\\

\begin{figure}[tp]
	\begin{center}
		\includegraphics[scale=1]{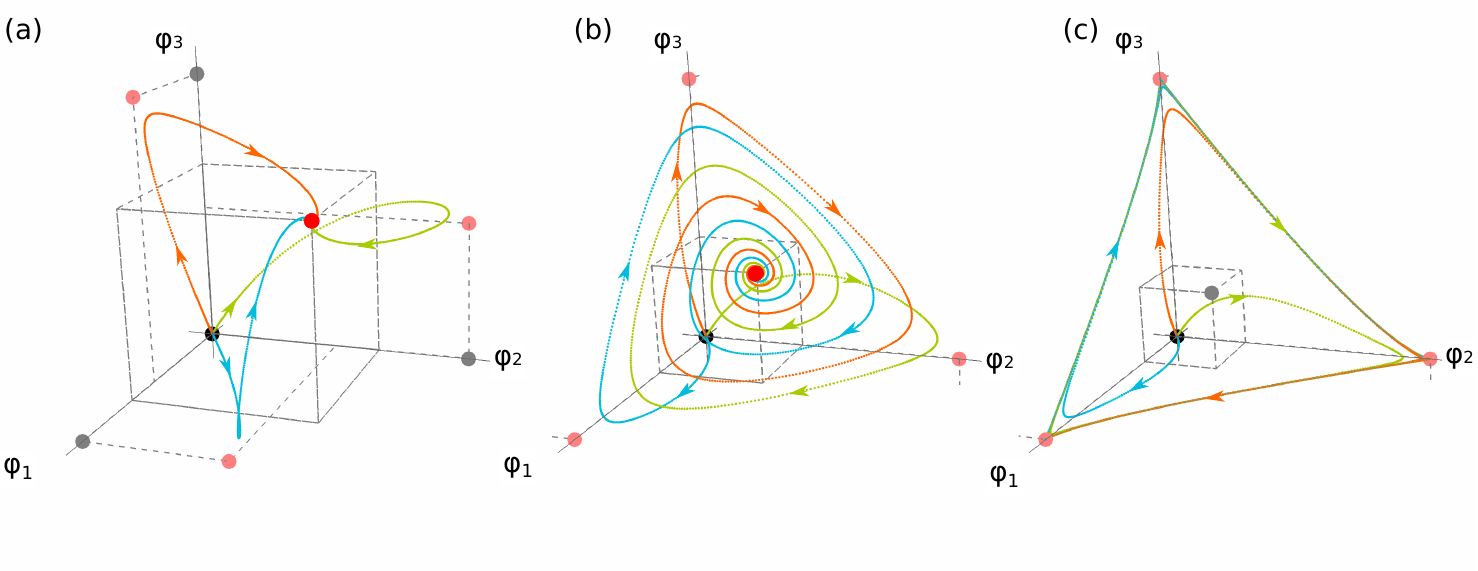}
		\end{center}
	\caption{Fixed point and trajectories of a (3,1)-game for different choices of parameters. Dots represent fixed points, unstable (black), stable (red), saddle with two stable directions (pink), and saddle with one stable direction (gray). Different colors of curves present slightly different initial conditions, all in the vicinity of an unstable fixed point (zero fixed point). (a) $\gamma>\kappa$ trajectories first get attracted by a saddle fixed point (pink), since there is a stable direction from a zero fixed point to saddles, and then get repelled to the stable coexistence fixed point. (b) The  coexistence fixed point for $\kappa/2<\gamma<\kappa$ is still stable, but the two saddles collided and exchanged stability, now the trajectories first go close to the saddles on the axes and then spiral into the stable fixed point. The existence of the oscillatory regime is indicated by the spiralling nature of the trajectories in the stable fixed point regime. (c)  The coexistence-fixed point for $\gamma<\kappa/2$ is unstable, the system evolves to a stable heteroclinic cycle. Before the trajectories evolve to the limit cycle, they approach the coexistence-fixed point, which indicates a stable direction from the unstable zero fixed to the saddle coexistence-fixed point.}\label{(3,1)traj}
\end{figure}

%\vskip5pt
If we now include the spatial component, the Jacobian $J$ gets an additional term $-\delta k^2$ on the diagonal, so that for $k\neq0$ the eigenvalues change, and so do the conditions for the stability of the fixed points. We can see that all eigenvalues are shifted by the value of $-\delta k^2$.
The first obvious consequence is that the zero-fixed point becomes stable for $\rho<\delta k^2$, a case, which is not of interest in view of applications. In order to have stable solutions, which do not cause the extinction of all species, we are interested in the case $\rho>\delta k^2$. The coexistence fixed point still has one stable direction for any choice of positive parameters, and the Hopf bifurcation happens for $\kappa/\gamma=2(\rho+\delta k^2)/(\rho-2\delta k^2)$, which is larger than 2 in the case of using all  parameters positive, so the stability regime increases.\\
Fixed points, which were ``exchanging their stability" by collision in a transcritical bifurcation, do no longer undergo  the bifurcation at the same point. The fixed point, corresponding to the survival of one species, changes its stability in one direction for $\kappa/\gamma=(\rho-\delta k^2)$, while the one with two species surviving goes through the transcritical bifurcation for $\kappa/\gamma=(\rho+\delta k^2)$. The dominant effect of the spatial dependence to the local stability of the fixed points is an extension of the parameter interval, in which the fixed points, or some directions, are stable.\\
As a solution of the system~(\ref{eq:MF(3,1)}) we observe no patterns in the regime, where the coexistence fixed point is stable, every site of the lattice is well mixed, every species evolves to the same value of the coexistence fixed point. Once we change the parameters such that this fixed point becomes unstable, the patterns observed on the lattice are spiral waves. These are oscillations both in space and time. Our linear stability analysis predicts oscillations in time, since the fixed point goes through a supercritical Hopf bifurcation, where a stable limit cycle is born.  Contrary to similar models of rock-paper-scissors games as considered in \cite{reichen1}-\cite{reichen6}, we do not choose a restriction on occupation numbers at sites, a feature that we had to compensate with a deletion reaction. At the end, however, this results in mean-field equations, for which one fixed point undergoes a supercritical Hopf bifurcation, without the need for introducing mutation reactions as in~\cite{mobilia}. %This allows for the derivation of amplitude equations around the subcritical Hopf bifurcations that will be considered in part II of this paper.

In order to compare how well the mean-field equations describe our original stochastic system, we compare results of the numerical solutions of the mean field partial differential equations~\ref{eq:MF(3,1)} with Gillespie simulations in both the fixed-point and the oscillatory regime, first for the (3,1)-game Eq.~\ref{eq:MF(3,1)}, and later for the (3,2) and (6,3)-games.

\subsection{Numerical solution for the(3,1)-game}
At $\kappa = 2$ (for $\gamma = 1$) the coexistence fixed point ($a^*=(\rho/(\gamma+\kappa),\rho/(\gamma+\kappa),\rho/(\gamma+\kappa))$) goes through a supercritical Hopf bifurcation. Due to this bifurcation a saddle limit cycle is created, one of the three eigenvalues stays negative, so there is a stable direction in the vector field towards the coexistence fixed point. At the same critical  value heteroclinic cycle connecting three one species saddles becomes stable. So in the oscillatory regime the system evolves to the heteroclinic cycle. If $\kappa$ is sufficiently close to the bifurcation point, the system will feel for a transient time the saddle-limit cycle, before it evolves to the heteroclinic cycle. For $\kappa<2$ the coexistence fixed point is stable, and independently of the initial conditions, the system will evolve to it.

We determine the numerical solutions of the mean field equations of a (3,1)-game for parameters $\rho=\gamma=1$, $\delta/dx=0.1$, and $\kappa=0.5$ (fixed point) and $4.0$ (oscillatory) regime, where $dx=2^{-7}$ is chosen as numerical grid constant and, if not otherwise stated, the grid size is 64 in units of dx, or 1 in dimensionless units.
\vskip3pt
{\bf Fixed point regime.} A plot in one dimension of the occupation of all sites x as a function of time t shows that the area of the $(x,t)$-diagram is gray (therefore not displayed) by using the following color scheme, which we also use later in other figures for the (3,1)-game. Each (x,t)-square in a lattice is  represented by a rgb-color scheme (r,g,b), where numbers r,g,b $\in$(0,1) are the value of species one, two, three, respectively, divided by the sum of values of all three species. This way species 1 is represented by red, species 2 by green, and species 3 by blue. Therefore the gray color corresponds to a well mixed distribution of all species through the lattice. In order to show that the gray color actually corresponds to the situation, where the concentrations at all sites approach the fixed-point value, we have checked that the concentrations $s_i, i=1,2,3$ at any site approaches the fixed-point value (it actually does so within less than 20 time units), and, vice versa, that after 100 time units the three concentrations at all sites have approached the values of the coexistence-fixed point.
\vskip3pt
{\bf Oscillatory regime.}
\begin{figure}[tp]
	\begin{center}
		\includegraphics[scale=1]{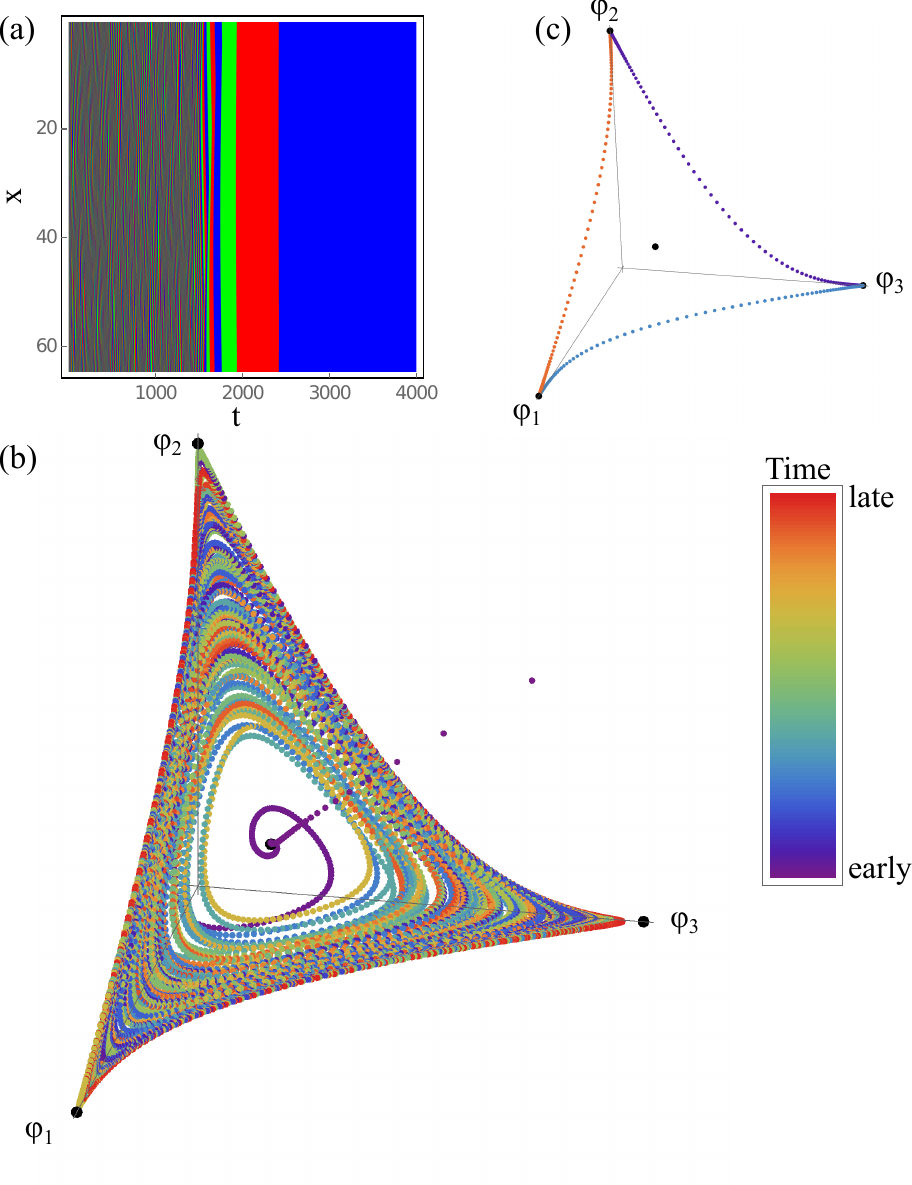}		
		\end{center}
	\caption{Spatio-temporal patterns (a), and phase portrait (b) and (c) of a (3,1)-game in an oscillatory regime at $\kappa=4$, far off the bifurcation point.Evolution in the phase space is represented by colored points at early (b) and late (c) time. Color of the points represents time. (b) evolution at lattice site 32 from 0-1700 t.u., (c) 1700-3000 t.u.}\label{(3,1)_array_beg}
\end{figure}

In figure\ref{(3,1)_array_beg} we show the spatial evolution in one dimension with periodic boundary conditions of a (3,1)-game in the oscillatory regime at $\kappa=4$, the other parameters are chosen as in the fixed-point regime. Here the system is far off the bifurcation point. Part (a) displays the spatio-temporal patterns until the system gets absorbed by one of the 1-species saddles, with the color scheme as indicated before. Part (b) shows the phase portrait for the first 1700 t.u. at the middle of the chain. Each dot represents a triple $(\varphi_1, \varphi_2,\varphi_3)$, where $\varphi_i$ denotes the concentration of the species $i$. The color of the dots represents the time instant upon time evolution, see the color bar in figure~\ref{(3,1)_array_beg}.  Large black dots represent the saddle-fixed points, three in the corners of the box are saddle-fixed points $FP_3-FP_5$, which are connected by the heteroclinic cycle, while the fixed point in the middle is the coexistence-fixed point $FP_1$ after the lost of stability. As transient behavior the system oscillates between all four saddle-fixed points, feeling also the saddle-limit cycle that is created by the Hopf bifurcation, even for $\kappa$ far away from the bifurcation value. After sufficiently long time, the system will evolve to a heteroclinic cycle, which is shown in the phase portrait of  figure~\ref{(3,1)_array_beg}(c).\\

In figure~\ref{(3,1)_tx_beg} we show the evolution in time at a site in the middle of the lattice, as displayed in figure~\ref{(3,1)_array_beg}. The coexistence-fixed point is at 0.25, indicated as the horizontal line at the fixed-point value. Trajectories oscillate between zero and one, which corresponds to the heteroclinic cycle, but are also attracted in irregular intervals by the small-amplitude limit cycle around the coexistence-fixed point, since the coexistence-fixed point became a saddle at the supercritical Hopf bifurcation point. As we have seen also for the (6,3)-game of the main text, the time evolution of $\varphi_i$ shows for later times the increasing dwell time of the trajectories in the vicinity of the saddles. It should, however,  be mentioned that the numerical integration on larger lattices reveals also regular large-amplitude oscillations between the three 1-species saddles for certain initial conditions, which are not characteristic for a stable heteroclinic cycle, for which the period of oscillations increases with every revolution. Since our stability analysis does not predict this kind of attractor, it is most likely transient behavior that seems to depend on the algorithm used for the numerical integration of the mean-field equations. The figure, corresponding to \ref{(3,1)_array_beg} (c) in Gillespie simulations, is figure~\ref{(3,1)1D_2} (b).

\begin{figure}[tp]
	\begin{center}
		\includegraphics[width=1\textwidth]{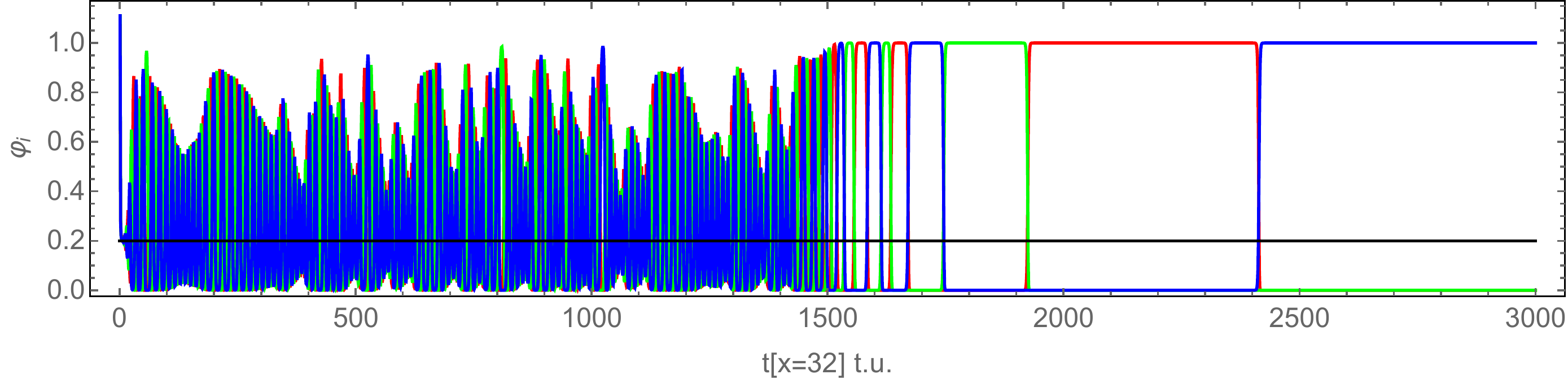}		
		\end{center}
	\caption{Concentration trajectories of the (3,1)-game corresponding to figure~\ref{(3,1)_array_beg} as a function of time at lattice site 32. System oscillates between a small saddle limit cycle surrounding the coexistence fixed point and the one species saddles until it reaches the heteroclinic cycle and eventually gets absorbed by one of the saddles due to numerical accuracy.}\label{(3,1)_tx_beg}
\end{figure}

\section{Stability analysis and numerical integration for the (3,2)-game}
\subsection{Stability analysis for the (3,2)-game}
The $(3,2)$-game is given by the set of equations
\begin{eqnarray}\label{eq:MF(3,2)}
\frac{\partial\varphi_1}{\partial t} & = & \nabla^2\varphi_1 + \rho\varphi_1 - \gamma\varphi_1^2 - \kappa\varphi_2\varphi_1 - \kappa\varphi_3\varphi_1 \nonumber \\
\frac{\partial\varphi_2}{\partial t} & = & \nabla^2\varphi_2 + \rho\varphi_2 - \gamma\varphi_2^2 - \kappa\varphi_1\varphi_2 - \kappa\varphi_3\varphi_2 \nonumber \\
\frac{\partial\varphi_3}{\partial t} & = & \nabla^2\varphi_3 + \rho\varphi_3 - \gamma\varphi_3^2 - \kappa\varphi_1\varphi_3 - \kappa\varphi_2\varphi_3 ,
\end{eqnarray}

with Jacobian
\begin{equation}
	J=
	\begin{pmatrix}
		\rho-2\gamma\varphi_1^*-\kappa(\varphi_2^*+\varphi_3^*) & -\kappa\varphi_1^* & -\kappa\varphi_1^*\\
		-\kappa\varphi_2^* & \rho-2\gamma\varphi_2^*-\kappa(\varphi_1^*+\varphi_3^*) & -\kappa\varphi_2^* \\
		-\kappa\varphi_3^* & -\kappa\varphi_3^* & \rho-2\gamma\varphi_3^*-\kappa(\varphi_1^*+\varphi_2^*)
	\end{pmatrix}.
\end{equation}

The fixed points and eigenvalues without and with diffusion term are given by the following table:

\vspace{5mm}
{\footnotesize
\centering
\begin{tabular}{c|c|c|c}
&	Fixed points & Eigenvalues for $D_\alpha=0$&  Eigenvalues for $D_\alpha\neq 0$ \\ \hline
$FP_1$ & (0, 0, 0) & ($\rho$, $\rho$, $\rho$) & ($\rho-\delta k^2$, $\rho-\delta k^2$, $\rho-\delta k^2$)\\ \hline
$FP_2$ & ($\frac{\rho}{\gamma+2\kappa}$, $\frac{\rho}{\gamma+2\kappa}$, $\frac{\rho}{\gamma+2\kappa}$) & ($-\rho$, $-\rho\frac{\gamma-\kappa}{\gamma+2\kappa}$, $-\rho\frac{\gamma-\kappa}{\gamma+2\kappa}$) &  ($-\rho-\delta k^2$, $-\frac{\rho(\gamma-\kappa)}{\gamma+2\kappa}-\delta k^2$, $-\frac{\rho(\gamma-\kappa)}{\gamma+2\kappa}-\delta k^2$)\\ \hline
$FP_3$ & ($\frac{\rho}{\gamma}$, 0, 0) & &\\
$FP_4$ & (0, $\frac{\rho}{\gamma}$, 0) & ($-\rho$, $\rho\frac{\gamma-\kappa}{\gamma}$, $\rho\frac{\gamma-\kappa}{\gamma}$)& ($-\rho-\delta k^2$, $\frac{\rho(\gamma-\kappa)}{\gamma}-\delta k^2$, $\frac{\rho(\gamma-\kappa)}{\gamma}-\delta k^2$)\\
$FP_5$ & (0, 0, $\frac{\rho}{\gamma}$) & &  \\\hline
$FP_6$ & (0, $\frac{\rho}{\gamma+\kappa}$, $\frac{\rho}{\gamma+\kappa}$) & &\\
$FP_7$ & ($\frac{\rho}{\gamma+\kappa}$, 0, $\frac{\rho}{\gamma+\kappa}$) & ($-\rho$, $\rho\frac{\gamma-\kappa}{\gamma+\kappa}$, $-\rho\frac{\gamma-\kappa}{\gamma+\kappa}$) & ($-\rho-\delta k^2$, $\rho\frac{\gamma-\kappa}{\gamma+\kappa}-\delta k^2$, $-\rho\frac{\gamma-\kappa}{\gamma+\kappa}-\delta k^2$)\\
$FP_8$ & ($\frac{\rho}{\gamma+\kappa}$, $\frac{\rho}{\gamma+\kappa}$, 0) & &
\end{tabular}.}
\vspace{5mm}

In the (3,2)-game, described by Eqs.~\ref{eq:MF(3,2)}, we have  eight different fixed points as in the (3,1)-case. None of the fixed points has any eigenvalues with imaginary part, in agreement with the numerical absence of oscillatory behavior, see Figs.~\ref{(3,2)st} and \ref{(3,2)ust} for typical trajectories of the system.

Here we also have an unstable zero-fixed point $FP_1$ with all three eigenvalues positive. The coexistence fixed point $FP_2$ is a stable focus for $\gamma>\kappa$. At $\gamma=\kappa$ two of the eigenvalues change sign. With increasing $\kappa/\gamma$ the fixed point becomes a saddle, having two unstable directions and one stable corresponding to the $-\rho$ eigenvalue. At the same time, the $FP_3$-$FP_5$  fixed points become stable, and the system evolves to one of these three fixed points, where only one species survives, depending on the initial conditions. The three remaining fixed points $FP_6$-$FP_8$ are always saddles, with two stable and one unstable direction. At $\gamma=\kappa$ these fixed points also undergo a bifurcation, in which two of the eigenvalues change sign, but without changing the number of stable and unstable directions, one of the stable directions, corresponding to the $-\rho\frac{\gamma-\kappa}{\gamma+\kappa}$ eigenvalue, becomes unstable, while an unstable direction corresponding to the eigenvalue $\rho\frac{\gamma-\kappa}{\gamma+\kappa}$ becomes stable. In summary, for $\gamma>\kappa$ the system evolves to a stable fixed point $FP_2$, where all species coexist, and for $\gamma<\kappa$ the system evolves to one of the three fixed  points $FP_3$-$FP_5$, where only one species survives. Which fixed point (species) this is, depends on the initial conditions.

\begin{figure}[tp]
	\begin{center}
		\includegraphics[scale=1]{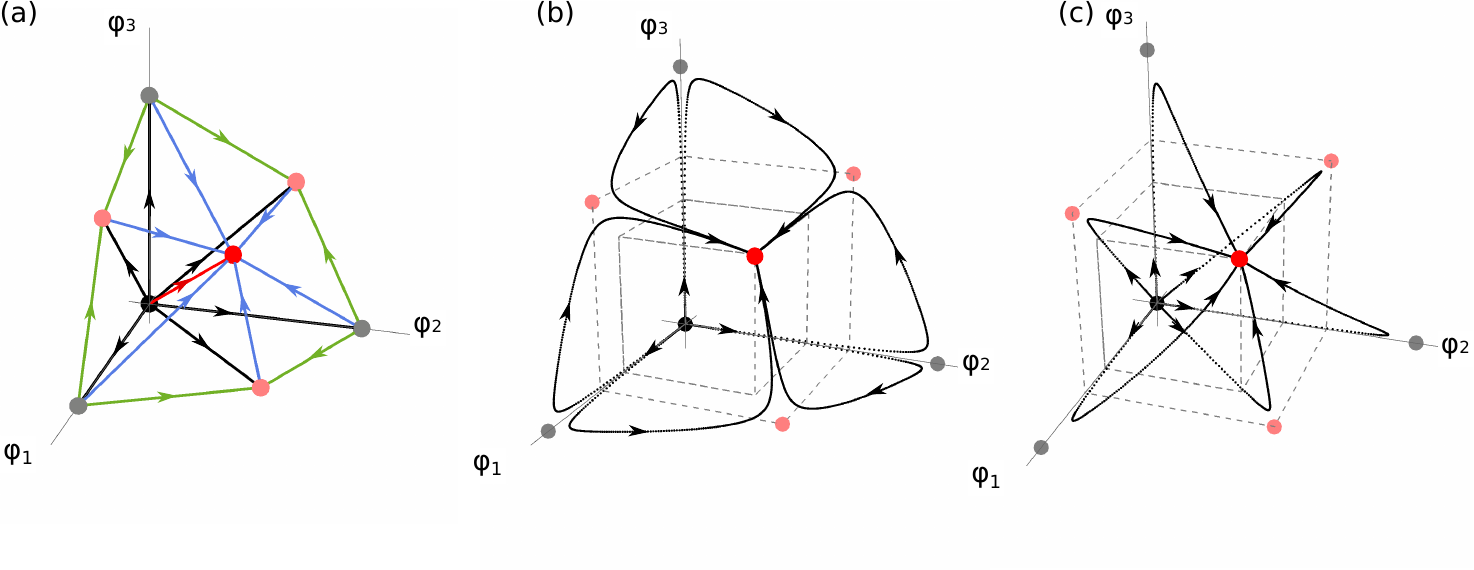}
		\end{center}
	\caption{Stability directions of a (3,2)-game between different fixed points (a) and trajectories (b)-(c) for different initial conditions, indicating the stability directions in a parameter regime, where the coexistence fixed point is stable $\gamma>\kappa$. (a) Arrows in lines point to the stability direction, green lines connect saddles with one stable (gray) and two stable (pink) directions, black lines the stability direction between the unstable fixed point (black) and saddles (gray and pink), the red line between the unstable and the stable fixed point, and blue lines between the stable and saddle fixed points. (b) and (c): Starting with slightly different initial conditions in the vicinity of the unstable fixed point, trajectories take different paths in (b) and (c): they first approach saddle fixed points, then following stable directions to finally approach the stable fixed point.}\label{(3,2)st}
\end{figure}

\begin{figure}[htbp]
	\begin{center}
		\includegraphics[scale=1]{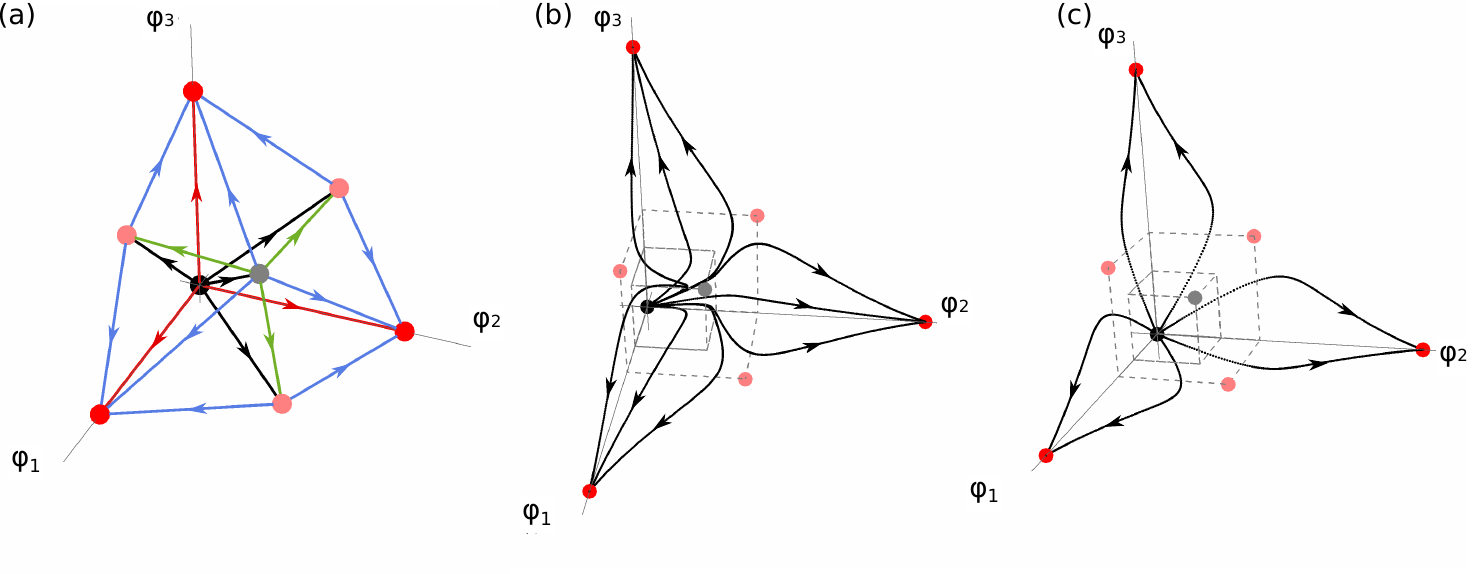}
		\end{center}
	\caption{Stability directions (a) and trajectories (b)-(c) of the (3,2)-game in the regime, where the coexistence fixed point is unstable $\gamma<\kappa$. The color code is the same as in figure~\ref{(3,2)st}. As the stability of the fixed points changes, so do the stable directions between the fixed points. The difference between (b) and (c) are the initial conditions.}\label{(3,2)ust}
\end{figure}
As in the (3,1) case, for $\delta\neq0$ the "zero"-fixed point $FP_1$ can become stable for $\rho<\delta k^2$ if $K\neq0$. The same happens with the fixed points $FP_6$-$FP_8$, which were always saddles in the case of $D_\alpha=0$. The parameter interval, in which other fixed points are stable, increases as in the case of the (3,1)-game. This leads to the possibility that all fixed points are stable for the same choice of parameters. %This does not apply to the single site, for which this scenario would not be possible,  but to the whole lattice.
On the lattice this may lead to the coexistence of different stable fixed points at different lattice sites. Which one is then approached at a single site depends on the initial conditions.
\vskip3pt
\subsection{Numerical solution of the (3,2)-game}
As we have seen, in  the (3,2)-game we have also two regimes of stationary behavior, but both regimes are fixed point regimes, for $\gamma=\rho=1$ the bifurcation point is at $\kappa=\gamma=1$. For $\kappa<1$ the system evolves to a coexistence-fixed point $a^*=(\rho/(\gamma+2\kappa), \rho/(\gamma+2\kappa), \rho/(\gamma+2\kappa))$. In this regime we see no spatio-temporal pattern formation. The system evolves fast into the fixed point.

\begin{figure}[htbp]
	\begin{center}
		\includegraphics[scale=0.85]{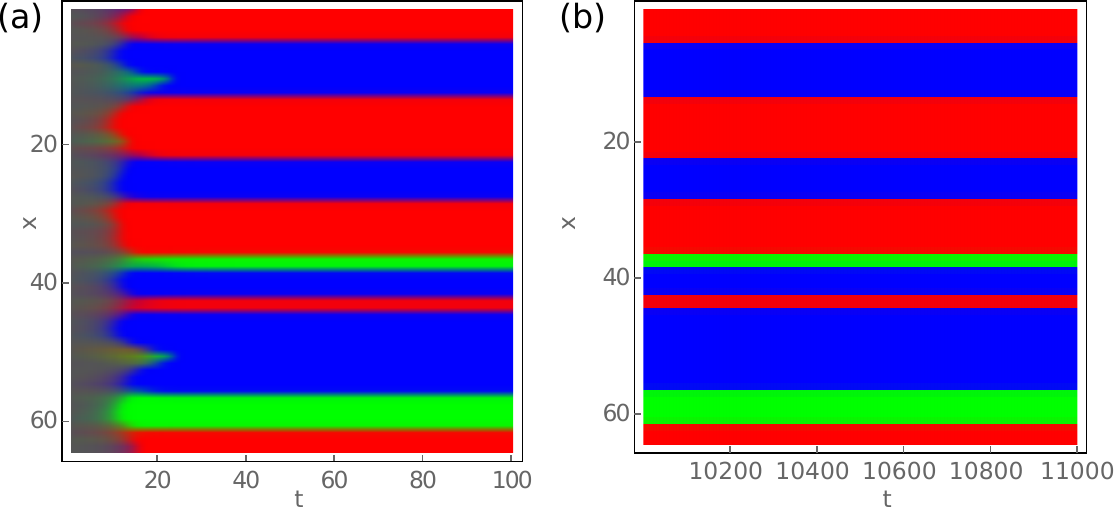}
		\end{center}
	\caption{Spatial evolution in one dimension of the (3,2)-game in a domain formation regime. The parameters are $\rho=\gamma=1$, $\delta/dx=0.1$, and $\kappa=4$, in the time interval (a) 0-100 t.u., and (b) 10000-11000 t.u. }\label{(3,2)_array_DOM}
\end{figure}

\begin{figure}[tp]
	\begin{center}
		\includegraphics[scale=0.9]{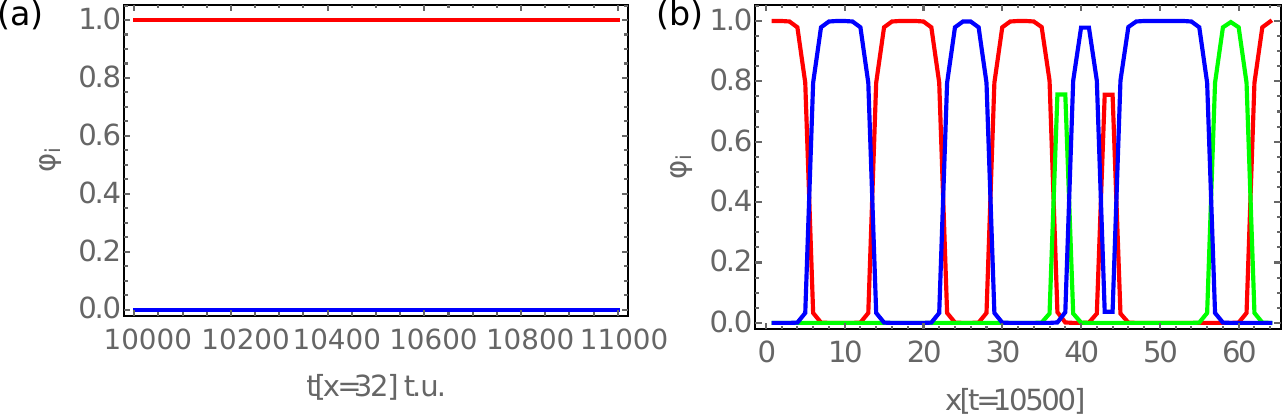}
		\end{center}
	\caption{Trajectories corresponding to figure~\ref{(3,2)_array_DOM}: (a) Time evolution of the concentrations at the middle (x=32) of the lattice towards a 1-species fixed point with one species dominant and two species at almost vanishing concentration, (b) spatial dependence at time 10500 t.u., showing the coexistence of domains.}\label{(3,2)_tx_DOM}
\end{figure}

For the above choice of parameters, it is for $\kappa>1$ that the coexistence fixed point becomes a saddle, with two eigenvalues positive (unstable directions), while three fixed points $a^*=(\rho/\gamma,0,0),(0, \rho/\gamma,0),(0,0, \rho/\gamma)$ become stable, so the system evolves into one of them, into which one depends on the initial conditions. For sufficiently large lattice size, or weak diffusion, the transient to one of the fixed points is very long in comparison to the transient to the coexistence-fixed point. So we distinguish between the coexistence-fixed point regime and the domain-formation regime.

\noindent {\bf Coexistence-fixed point regime.}
The results for this regime are not displayed, as no patterns are seen in the (x,t)-representation, while the concentrations  approach the fixed-point value rather fast for all x at given t and for all $t>5$ time units at given x.

\noindent {\bf Domain formation regime.}
The domain formation is displayed in figure\ref{(3,2)_array_DOM} at earlier (a) and later (b) times, where the domains have stabilized and the system is far off from the bifurcation point $(\kappa=4)$. For a sufficiently small diffusion  $\delta/dx=0.1$, the numerical solution of  Eq.~(\ref{eq:MF(3,2)}) seems to be stable in one dimension, we checked it up to time $10^7$ time units. Results of the corresponding Gillespie simulations are displayed in figure~\ref{(3,2)1D_1} (a)-(c) below.

%%%%%%%%%%%%%%%%%%%%%%%%%%%%%%%%%%%%%%%%%%%%%%%%%%%%%%%%%%%%%%%%%%%%%%%%%%%%%%%%%%%%%%%%%%%%%%%%%%%%%%%%%%%%%%%%%%%%%%%%%%%%%%%%
%%%%%%%%%%%%%%%%%%%%%%%%%%%%%%%%%%%%%%%%%%%%%%%%%%%%%%%%%%%

\section{Gillespie simulations of the(3,1)-game}\label{(3,1)Gill}
The patterns, which form in a $(3,1)$-game on a two-dimensional lattice, are multiple spirals. The time for the patterns to form depends on the diffusion, the system size, the reproduction rate with respect to predation and deletion events, and on the volume $V$. On the  one and two-dimensional lattices, the trajectories oscillate in  time, while the patterns that form in space depend on the diffusion strength. For the two-dimensional lattice we present results for two values of the diffusion strength.  For the one-dimensional lattice we analyze in addition the dependence on the ratio of $\kappa/\gamma$, which determines the distance from the bifurcation point in parameter space; this distance has also impact on the patterns. In the vicinity of the bifurcation points precursors of ``the other side" of the bifurcations are visible. The extension of the lattices that we show in the figures below are  $64\times 64$ in two-dimensions and $1\times64$ in one dimension.

\begin{figure}[htp]
	\begin{center}
		\includegraphics[scale=0.8]{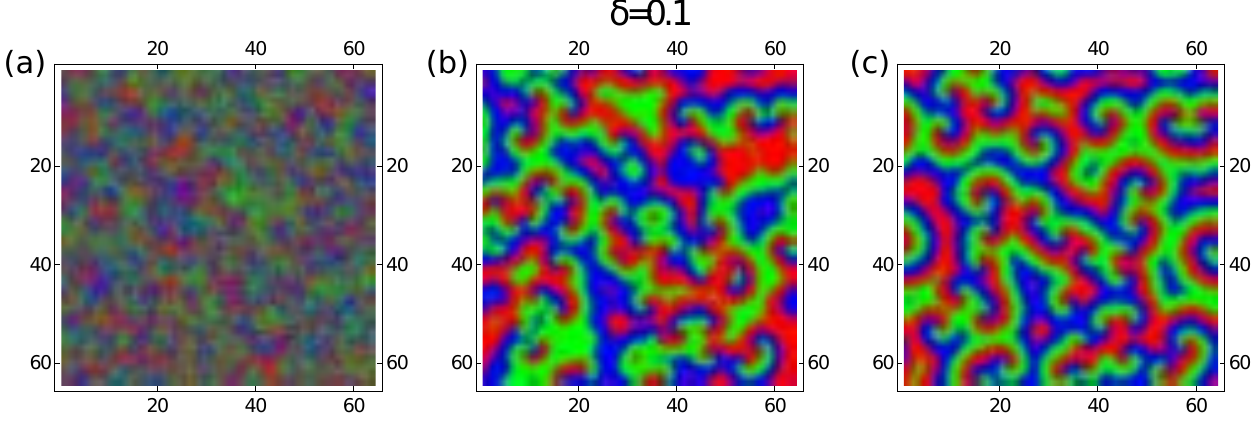}
	\end{center}
	\caption{Snap shots of the (3,1) game in the limit-cycle regime at $1000\cdot 2^{15}$ (a), $10000\cdot 2^{15}$ (b), and $30000\cdot 2^{15}$ (c) GS. The patterns formed on the lattice are multiple spirals. The parameter values are $\rho= 1$, $\kappa=1$, $\gamma=0.2$, and $\delta=0.1$, corresponding to slow diffusion.}\label{(3,1)2D_1}
\end{figure}

\begin{figure}[htp]
	\begin{center}
		\includegraphics[scale=0.8]{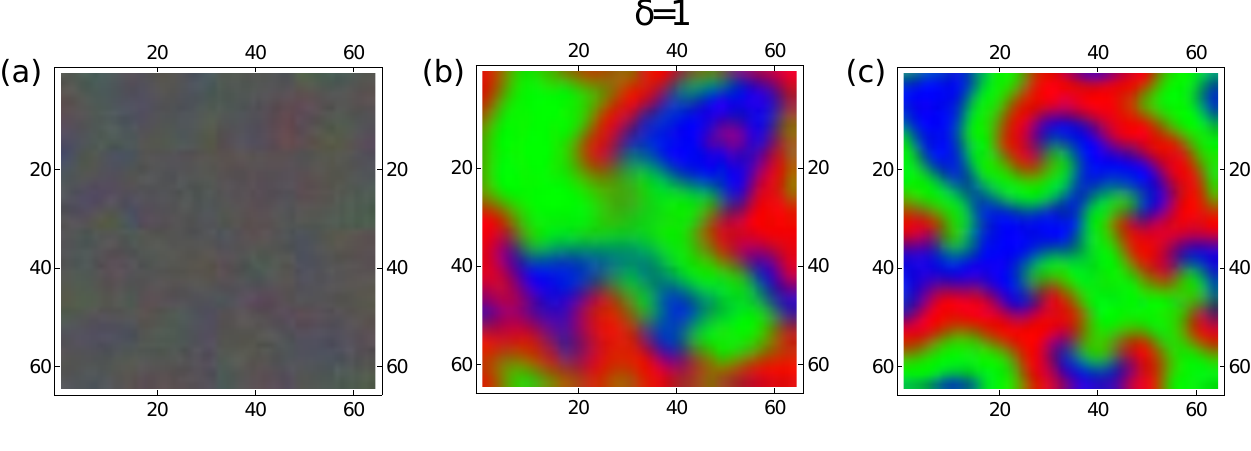}
		\end{center}
	\caption{Snap shots of the (3,1)-game for the same set of parameters as in figure~\ref{(3,1)2D_1}, but for faster diffusion, $\delta=1$. Snap shots are taken at times $1000\cdot 2^{15}$, $10000\cdot 2^{15}$, and $30000\cdot 2^{15}$ GS. A smaller number of spirals form with wider spiral arms than for slower diffusion.}\label{(3,1)2D_2}
\end{figure}

\begin{figure}[htbp]
	\begin{center}
		\includegraphics[scale=0.8]{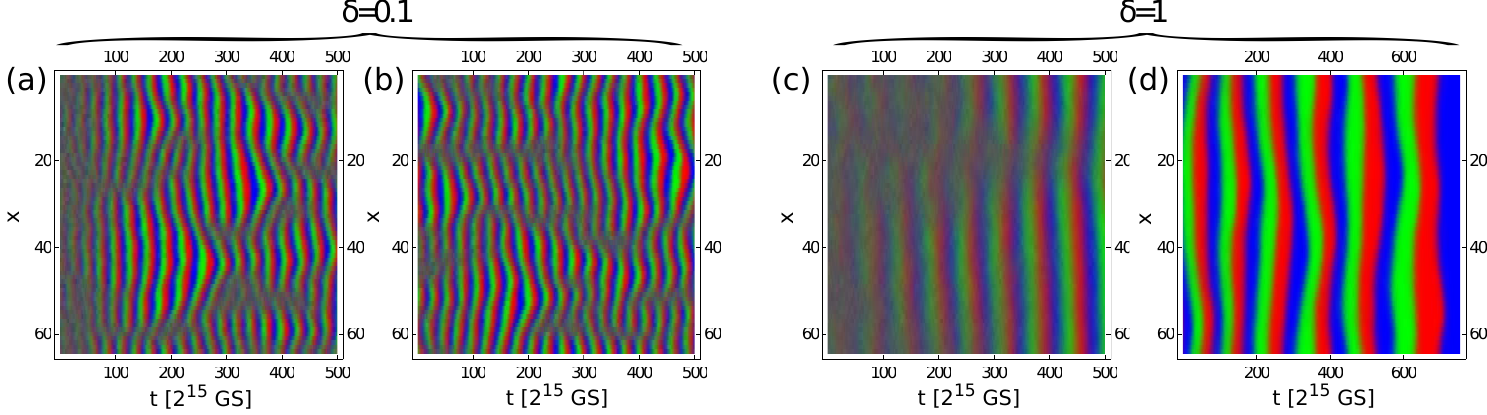}
		\end{center}
	\caption{Pattern formation in the (3,1)-game on a one-dimensional lattice, for a parameter regime close to the bifurcation point $\kappa=1.1$ and for two diffusion strengths $\delta=0.1$ for (a) and (b), and $\delta=1$ for (c) and (d). The time evolution on the lattice is shown for a period  of $500\cdot2^{15}$ GS, starting at time 0 (a), $30000\cdot2^{15}$ (b),  $0\cdot2^{15}$ (c) GS. Panel (d) shows an evolution from $(4000-4700)\cdot2^{15}$ GS. Other parameters are $\rho =\gamma=0.5$. For further comments see the text.}\label{(3,1)1D_1}
\end{figure}

\begin{figure}[h!]
	\begin{center}
		\includegraphics[scale=0.8]{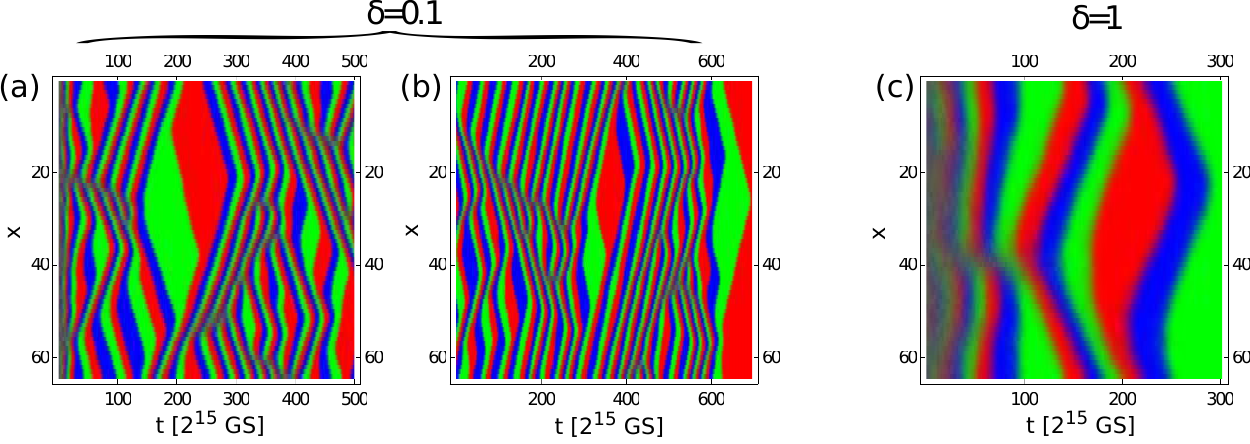}
		\end{center}
	\caption{Pattern formation in the (3,1)-game on a one-dimensional lattice as in figure~\ref{(3,1)1D_1}, but for $\kappa=2$ far off the bifurcation point. Panels (a) and (b) show the evolution of patterns for weak diffusion $\delta/dx=0.1$, (c) and (d) for strong diffusion $\delta=1$.  Extinction events happen for both strengths of diffusion, around $15700\cdot2^{15}$ GS for weak diffusion and around $300\cdot2^{15}$ GS for strong diffusion. Panel (a) shows the evolution in space for the time interval $0-500\cdot2^{15}$ GS, panel (b) for $(15000-15700)\cdot2^{15}$ GS, and (c) for $0-300\cdot2^{15}$ GS.}\label{(3,1)1D_2}
\end{figure}

Both on the one-dimensional and two-dimensional lattices, stronger diffusion causes wider patterns with respect to the system size, as can be seen from the comparison of the last panels of figures~\ref{(3,1)2D_1} and \ref{(3,1)2D_2}, figure~\ref{(3,1)1D_1} (b) and ~\ref{(3,1)1D_1} (d), and figures~\ref{(3,1)1D_2} (b) and \ref{(3,1)1D_2} (c). In case of weak diffusion none of the species goes extinct for a period of $2^{30}$ GS, while for stronger diffusion an extinction of two species (green and red) happens already at $4700\cdot2^{15}$ GS. The patterns look qualitatively similar: almost vertical waves that propagate in space and time. Figure~\ref{(3,1)1D_2} (b) should be compared with the mean-field prediction of figure~\ref{(3,1)_array_beg} (c), although the $\kappa$-values were chosen differently, but out of the same regime. The patterns qualitatively agree.

For stronger diffusion the spatial extension of waves is wider, the period of oscillations is larger,  they are more homogeneous on the lattice at a given instant of time.  As expected, the stronger the diffusion, the more homogeneous the lattice looks like.

Stronger diffusion also leads to a faster extinction of all but one species, this can be explained in the following way. In the stochastic simulations, extinctions happen after some time, and since the reproduction rate is proportional to the existing individuals of a certain kind, no ``resurrection" is possible. Extinctions are caused both by the finite spatial size and by the finite number of individuals. The finite size goes along with a smaller total number of individuals on the grid, while the finite number of individuals per site is tuned by the parameter V. Whenever the number of individuals is smaller, the  size of the demographic fluctuations gets larger in relation to average occupations of the deterministic limit. These fluctuations in the trajectories in phase space can kick the system into a parameter regime, where only a subset of species survives as the final fate of the system; the kick happens the faster, the stronger the fluctuations. Also for stronger diffusion, extinction happens faster than for weak diffusion, since the system  faster feels the finite grid size. Since extinction events do not happen coherently all over the grid, extinction of a certain species all over the grid takes the longer, the  larger the grid.

One may wonder whether strong diffusion can compensate for a large grid size with smaller demographic fluctuations and accelerate the extinction of species. For the range of parameter values, which we have chosen, the effect of demographic fluctuations dominated the impact of diffusion, so that extinction events decreased with increasing system size for both weak and stronger diffusion.

Figure~\ref{(3,1)1D_1} (d) shows the time evolution on a one-dimensional lattice for strong diffusion and close to the bifurcation point. This regime allows for more regular oscillations in time (compare figure~\ref{(3,1)1D_1} (d) with ~\ref{(3,1)1D_2} (c))  and a homogeneous arrangement in space.
It is the distance from the bifurcation point in the parameter space influences the patterns. The larger the distance from the bifurcation point is, that is, the larger $\kappa/\gamma$ ratio, the more disordered the patterns are,  see figure~\ref{(3,1)1D_2} (c). The reason is that the amplitudes of the ``noisy limit cycles'' \footnote{By `` noisy limit cycle" we mean the counterpart of the stochastic trajectory to the limit cycle in the deterministic limit.} are the larger, the farther the Hopf bifurcation  to the stable fixed point is. The larger their amplitude, the closer the trajectories come to the unstable fixed points, which are saddles, so the trajectories first get attracted and later repelled by the saddles and this way distorted, as we see  in figure~\ref{(3,1)1D_2} (c).  Although the width of the wave-like patterns, which are oscillations  in space and time, are of the same order for both considered ratios of $\kappa/\gamma$, for a larger ratio of $\kappa/\gamma$ this width fluctuates more, see figures~\ref{(3,1)1D_1} and \ref{(3,1)1D_2}. The "V"- and "Z"-like shapes of waves on an $(x,t)$-lattice are analogous to the multiple spiral centers on a two-dimensional lattice, in 2D we find multiple sources, which emit wave fronts that propagate in (t,x)-space and create "V" and "Z" shapes, when colliding with each other.

%\FloatBarrier

\section{Gillespie simulations of the (3,2)-game}\label{(3,2)Gill}
\begin{figure}[tp]
	\begin{center}
		\includegraphics[scale=0.8]{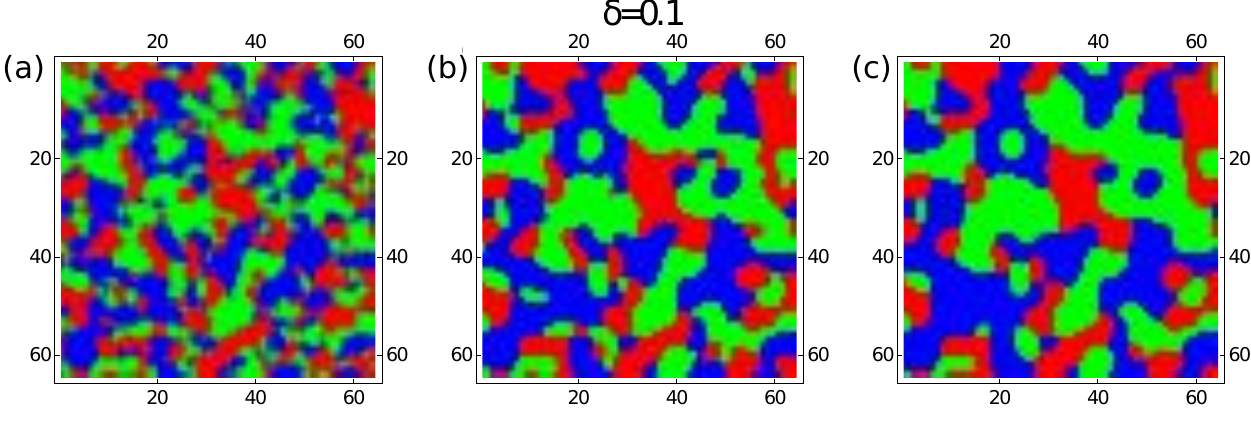}
		\end{center}
	\caption{Snap shots of the (3,2)-game on a two-dimensional lattice show domain formation. The parameters are $\kappa=\rho=1$, $\gamma=0.2$, and $\delta=0.1$, corresponding to slow diffusion. Snap shots are taken at $1000\cdot2^{15}$, $10000\cdot2^{15}$, and $30000\cdot2^{15}$ GS.}\label{(3,2)2D_1}
\end{figure}

\begin{figure}[tp]
	\begin{center}
		\includegraphics[scale=0.8]{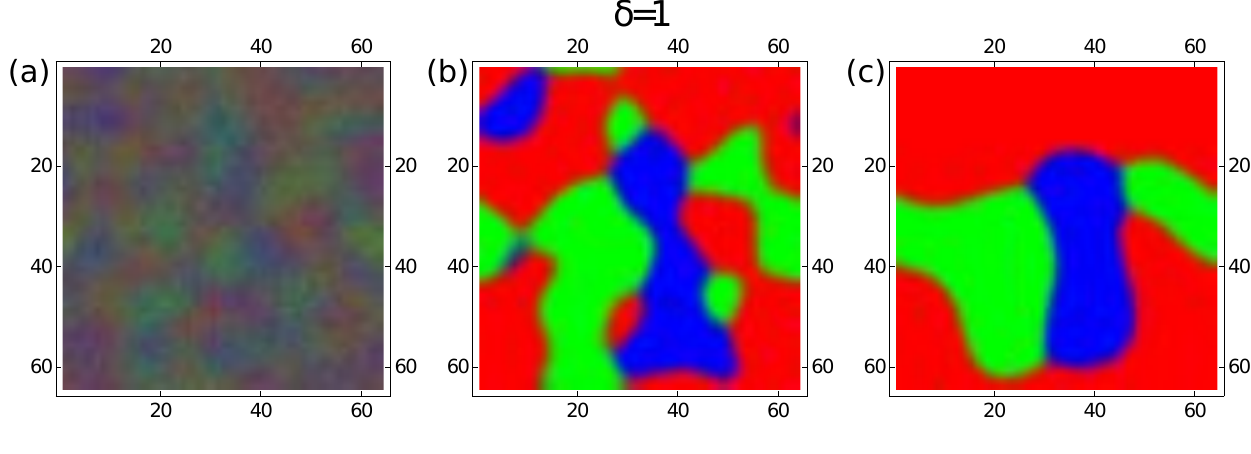}
		\end{center}
	\caption{Pattern formation in the (3,2)-game as in figure~\ref{(3,2)2D_1}, but for strong diffusion $\delta=1$. Snap shots are taken at $1000\cdot2^{15}$, $5000\cdot2^{15}$, and $30000\cdot2^{15}$ GS.}\label{(3,2)2D_2}
\end{figure}

The diffusion strength and the distance from the bifurcation point have a similar influence on the patterns as in the (3,1)-game. Figures~\ref{(3,2)2D_1} and \ref{(3,2)2D_2} show a domain formation in the $(3,2)$-game on a two-dimensional lattice for two different strengths of the diffusion constant. A domain consists of connected lattice sites, on which one species is dominant. Consider the number of individuals of the dominant species relative to the total number  of individuals on a certain site within domain. This number is smaller on the edges of the domain than in the center of it. In the center it often happens that the dominant species is the only one occupying the sites in the center.
The interaction between domains, and therefore between species, happens only at the boundaries of the domains. The boundaries are distinguished as sites with a well mixed occupation, with individuals of two or sometimes all three species.

\begin{figure}[tp]
	\begin{center}
		\includegraphics[scale=0.8]{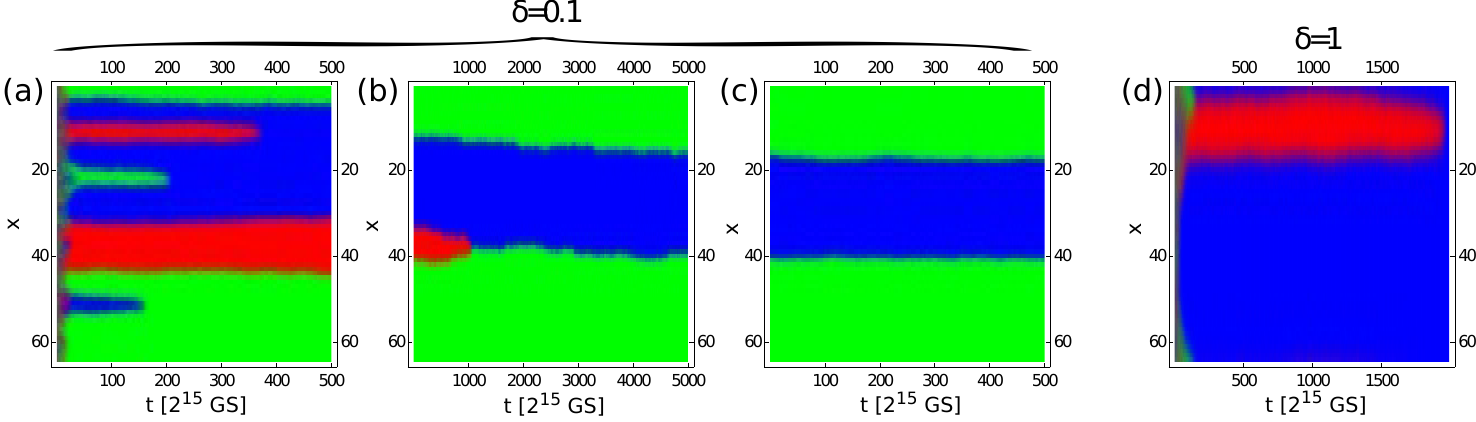}
		\end{center}
	\caption{Pattern formation for the (3,2)-game on a one-dimensional lattice for parameters $\rho=\gamma=0.5$, $\kappa = 1$, that is far off the bifurcation point, and $\delta$=0.1 (for (a), (b) and (c)), and $\delta=1$ (for (d)).  The panels show the evolution of patterns staring at 0 (a), $20000\cdot2^{15}$ (b), $32000\cdot2^{15}$ (c), and  0 (d) GS.}\label{(3,2)1D_1}
\end{figure}

Here the distance from the bifurcation parameter does not seem to influence the extinction times on one-dimensional lattices. The corresponding figures can be found in \cite{darkathesis}.
The time it takes for one or more species to go extinct, is of the same order of magnitude  for different ratios of $\kappa/\gamma$ if the diffusion constant is the same. This is qualitatively different from the (3,1)-game, where the extinction times depend on the distance from the bifurcation point in parameter space. In case of the (3,1)-game, the mean-field solutions are a heteroclinic cycle in the infinite volume limit, in which all species coexist, so the main mechanisms by which extinction can happen in a  stochastic realization are demographic fluctuations and the finite lattice size. Differently in  the (3,2)-game: After the 3-species coexistence-fixed point becomes unstable, one-species fixed points become stable, so we have an extinction of all but one species already on the mean-field level for all $\gamma<\kappa$, and the diffusion in the continuum homogenizes the individual fixed-point values to one and the same collective one. Before this happens in the stochastic realization, first domains form with different species occupying the sites, where the one-species fixed points on the sites may differ in the surviving species, neither need the extinction events happen on all sites simultaneously. The faster the random walk on the grid, the more it resembles a strong diffusion in the space-continuum, the more homogeneous the occupation becomes, so that all sites tend to evolve to the same fixed-point value that tells us which species survives. For weaker diffusion or slower random walk, domains may coexist for a long time until a single species occupies the entire lattice. For comparison with the mean-field predictions of figure~\ref{(3,2)_array_DOM} (a) and (b) we show figure~\ref{(3,2)1D_1} (a) and (b) with (c), respectively. Patterns forming on a (t,x)-grid are horizontal stripes, consisting of sites on which only one species dominates. On the boundaries of the stripes, sites are well mixed. With time, some of the stripes narrow and afterwards disappear, until the whole lattice is populated by only one species. As in the (3,1)-game, for weak diffusion no extinction of all but one species happens within $2^{30}$ GS, while for strong diffusion it happens after  $1700\cdot2^{15}$ GS. For further details on the Gillespie simulations of the (3,2)-game we refer to \cite{darkathesis}.


\section{References}
\begin{thebibliography}{100}
\bibitem{nowak} Nowak M A and Sigmund K, 2004, \textit{Science} \textbf{303}(6), 793.

\bibitem{may} May R, \textit{Stability and Complexity in Model Ecosystems}, (Princeton University Press, Princeton, NJ, 1974) 2nd ed.

\bibitem{levin} Levin S A, 1974 \textit{Am.Nat.} \textbf{108}, 207.

\bibitem{durret} Durrett R and Levin S A, 1994 \textit{Theor. Popul. Biol.} \textbf{46}, 363.

\bibitem{hassel} Hassel M P, Comins H N, and May R M, 1994 \textit{Nature} (London) \textbf{370}, 290.

\bibitem{szabo} Szab\'o G and  Fath G, 2007 \textit{Phys. Rep.} \textbf{446}(4), 97.

\bibitem{perc}Szolnoki A, Mobilia M, Jiang L L, Szczesny B, Rucklidge A M, and Perc M, 2016 \textit{J.R.Soc.Interface} \textbf{11}, 20140735.

\bibitem{reichen1} Reichenbach T, Mobilia M, and Frey E, 2006 \textit{Phys. Rev. E} \textbf{74}, 051907.

\bibitem{reichen2} Reichenbach T, Mobilia M, and Frey E, 2007 \textit{Phys. Rev. Lett.} \textbf{99}, 238105.

\bibitem{reichen3} Reichenbach T, Mobilia M, and Frey E, 2007 \textit{Nature Lett.} \textbf{448}, 1046.

\bibitem{reichen4} Reichenbach T and Frey E, 2008 \textit{Phys. Rev. Lett.} \textbf{101}, 058102.

\bibitem{reichen5} Reichenbach T, Mobilia M, and Frey E, 2008 \textit{J. Theor. Biol.} \textbf{254}, 368.

\bibitem{reichen6} Berr M, Reichenbach T, Schottenlohner M, and Frey E,  2009 \textit{Phys. Rev. Lett.} \textbf{102}, 048102.

\bibitem{goldenfeldwoese} Goldenfeld N and Woese C, 2011 \textit{Ann. Rev. Condens. Matter Phys.} {\bf 2}, 375.

\bibitem{durney} Durney C H, Case S O, Pleimling M, and Zia R K P, 2011 \textit{Phys. Rev. E} \textbf{83}, 051108.

\bibitem{case}  Case S O, Durney C H, Pleimling M, and Zia R K P, 2010 \textit{Europhys. Lett.} \textbf{92}, 58003.

\bibitem{28} Durney C H, Case S O, Pleimling M, and Zia R K P, 2012 \textit{J. Stat. Mech.} P06014.

\bibitem{arXiv:1205.4914} Roman A, Konrad D, and Pleimling M, 2012 \textit{J. Stat. Mech.} P07014.

\bibitem{szabosznaider} Szab\'o G, and Sznaider G A, 2004 \textit{Phys. Rev. E} \textbf{69}, 031911.

\bibitem{luetz} L\"utz A F, Risau-Gusman S, and Arenzon J J, 2013 \textit{J.Theor.Biol.} \textbf{317}, 286.

\bibitem{hua} Hua D-Y, Dai L-C and Lin C, 2013 \textit{Europhys. Lett.} \textbf{101}, 38004.

\bibitem{m1} Roman A, Dasgupta D, and Pleimling M, 2013 \textit{Phys. Rev. E} \textbf{87}, 032148.

\bibitem{m2} Mowlaei S, Roman A, and Pelimling M, 2014 \textit{J. Phys. A: Math. Theor.} \textbf{47}, 165001.

\bibitem{m3} Roman A, Dasgupta D, and Pleimling M, 2016 \textit{J. Theor. Biol.} \textbf{403}, 10.

\bibitem{josef} Hofbauer J, Kon R, and Saito Y, 2008 \textit{J. Math. Biol.} \textbf{57}, 863.
	
\bibitem{mobilia1}
	Szczesny B, Mobilia M, and Rucklidge A M, 2013 \textit{EPL} \textbf{102}, 28012.

\bibitem{mobilia2} Szczesny B, Mobilia M, and Rucklidge A M, 2014 \textit{Phys. Rev. E} \textbf{90}, 032704.

\bibitem{goldenbutler} Butler T, and Goldenfeld N, 2009 \textit{Phys. Rev. E} \textbf{80}, 030902(R).

\bibitem{darkathesis} Labavi\'c D, {\it Deterministic versus stochastic descriptions of nonlinear dynamical units}, (Jacobs University Bremen, Bremen 2015) PhD thesis.

\bibitem{gillespie}
	Gillespie D T, 1977 \textit{J. Phys. Chem.} \textbf{81}, 2340.

\bibitem{elf}
	Elf J and Ehrenberg M, 2004 \textit{Systems Biology} \textbf{1}(2).

\bibitem{vanKampen}
	Van Kampen N G, \textit{Stochastic Processes in Physics and Chemistry}, (Elsevier, Amsterdam, The Netherlands, 2005).

\bibitem{bookkuramoto} Kuramoto Y, textit{Chemical Oscillations, Waves, and Turbulence}, Springer Series in Synergetics \textbf{19},(Springer, Berlin, 1984).

\bibitem{cianci}
	Cianci C and Carletti T, 2014 \textit{Physica A} \textbf{410}, 66.


\end{thebibliography}

\section{References}
\begin{thebibliography}{100}

\bibitem{aubin} Aubin J P, and Sigmund K, 1988 \textit{J. Comput. Appl. Mathem.} \textbf{22}, 203.
    
\bibitem{hofbauer} Hofbauer J and Sigmund K, \textit{Evolutionary Games and Population Dynamics}, (Cambridge University Press, Cambridge, 1998).

\bibitem{reichen1} Reichenbach T, Mobilia M, and Frey E, 2006 \textit{Phys. Rev. E} \textbf{74} 051907.

\bibitem{reichen2} Reichenbach T, Mobilia M, and Frey E, 2007 \textit{Phys. Rev. Lett.} \textbf{99} 238105.

\bibitem{reichen3} Reichenbach T, Mobilia M, and Frey E, 2007 \textit{Nature Lett.} \textbf{448} 1046.

\bibitem{reichen4} Reichenbach T and Frey E, 2008 \textit{Phys. Rev. Lett.} \textbf{101} 058102.

\bibitem{reichen5} Reichenbach T, Mobilia M, and Frey E, 2008 \textit{J. Theor. Biol.} \textbf{254} 368.

\bibitem{reichen6} Berr M, Reichenbach T, Schottenlohner M, and Frey E,  2009 \textit{Phys. Rev. Lett.} \textbf{102} 048102.

\bibitem{mobilia}	Szczesny B, Mobilia M, and Rucklidge A M, 2013 \textit{EPL} \textbf{102} 28012.

\bibitem{darkathesis} Labavi\'c D, {\it Deterministic versus stochastic descriptions of nonlinear dynamical units}, (Jacobs University Bremen, Bremen, 2015) PhD thesis.






\end{thebibliography}
\end{document}